\documentclass{aa}  
\usepackage{graphicx}
\usepackage{txfonts}
\usepackage{newtxtext,newtxmath} 
\usepackage[T1]{fontenc}
\usepackage{ae,aecompl}
\usepackage{amsmath}	
\usepackage{amssymb}	
\usepackage{booktabs}
\usepackage{color}
\usepackage{seqsplit}
\usepackage{ulem}
\usepackage{hyperref}
\usepackage{appendix}
\usepackage[figuresright]{rotating}
\usepackage{graphicx}
\graphicspath{{plots/}}
\usepackage{flafter}

\newcommand{\orcid}[1]{\href{https://orcid.org/#1}{\protect\includegraphics[width=8pt]{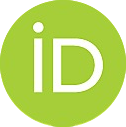}}}

\newcommand{\kms}{km\,s$^{-1}$}
\newcommand{\nstar}{N_{\rm s}}

\newcommand{\rvir}{r_{\rm vir}}

\newcommand{\amuse}{\texttt{AMUSE}}
\newcommand{\nb}{\texttt{NBODY6++GPU}}
\newcommand{\nbm}{\texttt{NBODY6++GPU-MASSLESS}}

\newcommand{\msun}{\,{\rm M}_\odot}

\newcommand{\ffc}{\texttt{ffc}}

\newcommand{\halfmass}{r_{\rm hm}}
\newcommand{\clustermass}{M_{\rm cl}}
\newcommand{\tcross}{t_{\rm cr}}
\newcommand{\trelax}{t_{\rm rlx}}

\newcommand{\fff}[1]{\textcolor{black}{ #1}}

\newcommand{\ffff}[1]{\textcolor{black}{ #1}}

\def\apjl{ApJ}          
\def\apjs{ApJS}          
\def\ao{Appl.~Opt.}          
\def\aap{A\&A}          
{}
{}
{}
{}
\begin{document} 

   \title{Dynamical evolution of massless particles in star clusters with \nbm}
   \titlerunning{Dynamical evolution of \ffc{}s}
   \authorrunning{Flammini Dotti et al}
   \subtitle{II. The long-term evolution of free-floating comets}

\author{Francesco Flammini Dotti\thanks{ff2415@nyu.edu}\inst{1,2,3,4}\orcid{0000-0002-8881-3078},
M.B.N. Kouwenhoven\inst{5}\orcid{0000-0002-1805-0570},
Kai Wu\inst{3}\orcid{0000-0003-0349-0079},
Abbas Askar\inst{6}\orcid{0000-0001-9688-3458},
Peter Berczik\inst{6,7,8}\orcid{0000-0003-4176-152X},
Mirek Giersz\inst{6}\orcid{0000-0002-5987-5077},
Rainer Spurzem\inst{3,9,10}\orcid{0000-0003-2264-7203},
Ian Dobbs-Dixon\inst{1,2}\orcid{0000-0002-4989-6501}
  }

  \institute{Department of Physics, New York University Abu Dhabi, PO Box 129188 Abu Dhabi, UAE
  \and Center for Astrophysics and Space Science (CASS), New York University Abu Dhabi, PO Box 129188, Abu Dhabi, UAE
  \and Astronomisches Rechen-Institut, Zentrum f\"ur Astronomie der Universität Heidelberg, M\"onchhofstrasse 12--14, 69120, Heidelberg, Germany
  \and
  Dipartimento di Fisica, Sapienza, Universit\'a di Roma, P.le Aldo Moro, 5, 00185 - Rome, Italy
  \and
  Department of Physics, School of Mathematics and Physics, Xi'an Jiaotong-Liverpool University, 111 Ren'ai Road, \\
  Suzhou Dushu Lake Science and Education Innovation District, Suzhou Industrial Park, Suzhou 215123, P.R. China
  \and
  Nicolaus Copernicus Astronomical Centre Polish Academy of Sciences, ul. Bartycka 18, 00-716 Warsaw, Poland
  \and
  Szechenyi Istvan University, Space Technology and Space Law Research Center, H-9026 Gyor, Egyetem ter 1. Hungary
  \and
  Main Astronomical Observatory, National Academy of Sciences of Ukraine, 27 Akademika Zabolotnoho St, 03143 Kyiv, Ukraine
  \and
  National Astronomical Observatories and Key Laboratory of Computational Astrophysics, Chinese Academy of Sciences, 20A Datun Rd., Chaoyang District, 100101, Beijing, China
  \and
  Kavli Institute for Astronomy and Astrophysics, Peking University, Yiheyuan Lu 5, Haidian Qu, 100871, Beijing, China
  }

   \date{Received ---; accepted ---}

 
   \date{Received --; accepted --}

  \abstract
  {
   Comets, asteroids, planetesimals, free-floating planets and brown dwarfs, are continuously injected into the intra-cluster environment after expulsion from their host planetary systems or binary system. The dynamics of large populations of such free-floating comets (\ffc{}s) in a star cluster environment is not yet fully understood. }
   {We investigate the dynamical evolution of comet populations in star clusters, and characterize \fff{the} kinematics and ejection rates of \ffc{} in a star cluster. Moreover, we \fff{determine whether a different initial energy distribution} affects the mass segregation of the less massive population.}
   {We carry out simulations using the $N$-body code \nbm{} \citep{flamminidotti2025}, which allows fast integration of star clusters that contain large numbers of massless particles, to characterize the dynamics of  populations of low-mass particles with sub-virial and super-virial distributions. }
   {Comets do not participate in the mass segregation process, similarly to planet-size objects, regardless of their initial energy distribution. The latter is slightly changing the whole dynamical evolution at the start of the simulation. We only observe an initial relaxation or collapse of the objects for super-virial and sub-virial ratios, respectively. The external regions of the \ffc{}s population tend to be pulled back in the cluster core at the end of the simulation, suggesting the gravitational pull of the stars is pulling them back \fff{in the core}. This \fff{phenomenon occurs at later times} if the system in virial equilibrium.
   \fff{Compared to less massive bodies, brown dwarfs experiences more mass segregation the inner regions tend to be more mixed with the stellar population.} }
   {}


   \maketitle
%

\section{Introduction}

Modern research of the dynamical evolution of comets began with~\cite{Oort:1950aa}.
From the statistical properties of observed cometary orbits, Oort postulated the existence of a distant reservoir of comets with an inner radius of 50\,000~au and an outer radius of 150\,000~au as the origin of all the ``new'' long-time period comets. This hypothetical cloud of comets is now known as the ``Oort cloud'' and it is widely believed to be the source of the comets with orbital periods longer than 200~years. Several hundreds of comets have been identified in the recent centuries. These constitute only a tiny fraction of the comet reservoir in the Solar system. As comets in the Oort Cloud are not directly detectable, it is difficult to accurately estimate their abundance. A meticulous analysis of the kinematic properties of known long-period comets, as well as numerical simulations, can provide reasonable estimates of the present-day composition of the Oort Cloud \citep[e.g.][and references therein]{Oort:1950aa, Everhart:1967aa, Everhart:1967ab, Weissman:1982aa, Weissman:1996aa, Wiegert:1999aa, Dones:2004aa, Francis:2005aa, Fouchard:2013aa, Shannon:2015aa, pz2021a,pz2021b, bazyey2022, raymond2024}. Although results vary, these studies generally suggest that the number of comets in the brighter than $H<10.9$~mag in the outer Oort cloud is of order $10^{11}-10^{12}$. Numerical studies and extrapolations to smaller masses suggest that the present-day Oort cloud has a total mass of $2-40$ Earth masses \citep[see, e.g,][]{Francis:2005aa, Morbidelli:2005aa}.

During the epoch of planet formation, a large fraction of the planetesimals experienced gravitational scattering events that led to the ejection into the outer Solar system. The majority of these comets obtained velocities beyond the local escape velocity and became free-floating comets in the Galactic field, while others obtained highly eccentric orbits in the Oort cloud. Following the period of rapid ejection of comets, the Solar system entered a more quiescent phase, where comets are occasionally removed following a gravitational interaction with a planet or a neighbouring star. It is estimated that, over \fff{the course of $\approx 20$ Myr}, close encounters with neighbouring stars in the Galactic field, and the Galactic tidal force itself, have resulted into the removal of roughly 10\% of the Oort Cloud into interstellar space \citep[e.g.,][]{Biermann:1978aa, Weissman:1980aa, Dybczynski:2002aa, Torres:2019aa}.

Given the ubiquity of exoplanets in the Galactic field, it is not unreasonable to assume that a similar process has also occurred during early evolution of exoplanet systems, and that the Milky Way is filled with comets and other free-floating debris. In fact, several studies claimed to have found evidence for so-called exo-comets in system such as $\beta$~Pictoris \citep[e.g.,][and references therein]{Welsh:2016aa} and KIC~8462852 \citep[e.g.,][and references therein]{Bodman:2016aa, Boyajian:2016aa}. Dynamical capture of these objects is rare \citep[e.g.,][]{Valtonen:1982aa, Torbett:1986aa}. Occasionally, free-floating debris may interact with a planetary system, get captured, and possibly even collide with a star or planet.

Star clusters are potential candidates for hosting significant populations of free-floating planets and comets. The most recent discovery of candidate free-floating planets, made by \cite{miretroig2023} in the Scorpius association, includes thousands of planet-like objects and brown dwarfs. Free-floating planets have not yet been discovered in star clusters, although several candidates and few confirmed free-floating planets have been found in associations \citep[for a large microlensing survey, check][]{miretroig2023}. High-density environments, such as globular clusters, pose significant challenges for detecting planetary systems, let alone rare occurrences such as stellar occultation by free-floating planets. In nearby and less dense environments, such as open clusters, it is easier to detect such occurrences. However, they have not yet been found, as this would require a dedicated survey, similar to the work of \cite{miretroig2023}. Nevertheless, several studies \citet{Cai:2019aa, wukai2023,kai2024} with modeled debris disks hint the existence of large free-floating comet populations in star clusters. Their numerical simulations suggest that, among comet-like particles engulfed by flyby stars, a significant fraction of them will eventually reside inside their birth cluster. Moreover, the inclusion of disk-accompanying planets will further increase this fraction. 

Close encounters between comet-hosting stars and other cluster members can result in the tidal stripping of comets from their host planetary systems. 
\fff{Both encounters and the tidal stripping from the comet-hosting star} are responsible for the injection of debris into star clusters \citep[e.g.,][]{Eggers:1997aa, Dones:2015aa, Brasser:2008ab, Brasser:2008ac, Brasser:2015aa}. As the velocity-at-infinity at which comets are ejected from their host planetary system is typically much smaller than the local escape velocity, these comets remain part of the system and participate in the further dynamical evolution of the cluster, until their escape from the cluster through ejection or evaporation \citep[e.g.,][]{Wang:2015ab}. Such free-floating comets are captured by other stars \citep{Kouwenhoven:2010aa, Perets:2012aa, Parker:2017aa, Kokaia2020} \fff{and may be captured for larger times \citep[e.g.,][]{pena2023}}, or even during the planet formation process \citep{Pfalzner2019}. The balance between a steady injection rate and escape of comets determines the total number of free-floating comets that is present in a star cluster at a given time.\fff{The number of free-floating substellar objects injected in the Galactic field is large. The number free-floating planets (with Jupiter mass) is estimated to be 1/4th of the number of stars the Galaxy \citep{Mroz:2017aa}. Moreover, up to $10^{12}$ exocomets are estimated to be present in the Galactic field for every star \citep{seligman2022}.}

Ejection and capture of comets in star clusters occurs at a much higher rate than in the Galactic field. In these crowded stellar environments, many comets are expelled from their host stars due to the high frequency of close stellar encounters \citep[e.g.,][and references therein]{Cai:2019aa, 2019MNRAS.490...21H, Veras2019}. The following three processes are responsible for the presence of free-floating comets in star clusters: (i) ejection of comets from their birth planetary systems due to interactions with (proto-)planets, (ii) ejections of comets from their planetary system following a close encounter with another star cluster member, and (iii) escape from their planetary system as their host star evolves and experiences phases of mass loss \citep[see, e.g.,][]{Veras:2012aa}. 

Free-floating comets ejected from planetary systems in star clusters most likely have comparable speeds, and these ejection speeds are typically smaller than the local velocity dispersion in a star cluster \citep[e.g.,][and references therein]{Spurzem:2009aa, Zheng:2015aa, 2019A&A...624A.110F, flamminiblack, wangyihan}. Consequently, free-floating comets can remain bound to the star cluster for many dynamical times prior to their escape from the cluster \citep[see][]{Wang:2015ab}, during which the possibility for re-capture of a comet by another star arises. 

Star cluster members occasionally pair up and form new binary systems. 
Such newly-formed systems are often transient \citep[e.g.,][]{Moeckel:2011aa}, although some may survive for much longer times \citep[e.g.,][]{Shu2020}, in particular when they reside in the outskirts of the star cluster or 
when they are in the process of escaping from the star cluster \citep{Kouwenhoven:2010aa}.  Free-floating planets and free-floating planetary debris can also be captured \citep{Perets:2012aa, parker2012, Zheng:2015aa,Portegies:2020aa}. Such captured bodies may be common \citep[e.g.,][]{Siraj:2020aa}, and can be identified through their orbital parameter distributions \citep[e.g.,][]{Siraj:2018aa}.

\fff{The process of mass segregation in different types of star clusters has been extensively studied intensively \citep[e.g.,][]{Fregeau2001_LightTracers,Allison2009_MST,Olczak2011_effMST,DeVita2019_CorrSegStruct,Adamo2020_StarClustersNearAndFar,Baumgardt2022_MSGlobulars}. The process depends on the stellar mass spectrum, and is affected by both two-body relaxation and stellar evolution. However, studies on the impact of mass segregation on brown dwarfs are rare. It is important to highlight the study of \cite{parker2016}, which distinguishes the concept of mass segregation from the process of energy equipartition.}

In this study we consider the dynamical evolution of a population of free-floating comets (\ffc{}s) in star clusters. We will analyse their kinematics using different energy distribution for the \ffc{} population. Our findings are applicable to any population of bodies can be approximated as massless, including comets, asteroids, planetesimals, and free-floating planets. 

This article is organised as follows. In Section~\ref{section:method} we briefly describe our numerical method and initial conditions. In Section~\ref{section:results3}, we discuss the dynamical evolution of the star clusters and the \ffc{} populations. We will also study the \ffc{} ejection rates. In Section~\ref{section:result4}, we try to theoretically predict whether the segregation of these small particle is possible or not. Finally, we summarise our results and  discuss the implications of our findings in Section~\ref{section:conclusions}.



\section{Methodology and initial conditions}\label{section:method}

\begin{table*}
	\centering
	\caption{Initial conditions of the star cluster models. Each cluster model is named based on the \fff{value of the virial ratio of the \ffc{} population}. The stellar component is the same in all models, only the \ffc{}s population changes energy distribution.
    }\label{table:details}	
	\resizebox{\textwidth}{!}{\begin{tabular}{llll} 
		\hline\hline
		Object type & Stars &Common features&  \ffc{} models with $Q= 0.25$, 0.5, and 0.75\\
		\hline
        \hline
        Model Name for each \ffc{}s subset && &  C025 \& C05 \& C075 \\
        \hline
		Number of objects & 10\,000 && 30\,000 \\
        Number of stars in binaries, $n_{\rm bin}$ & 1\,000 && \\ 
		Object initial mass function & \cite{2001MNRAS.322..231K},  $0.08-150~\msun$ && Equal mass, $m_{\rm ffc} = 10^{-11} \msun $\\
		Total mass, $M_{\rm cl}$  &   $6.62 \times  \ 10^3\,\msun$ && $3 \ \times 10^{-7} \msun $  \\   
		Density profile && \cite{Plummer:1911aa}  &  \\
        Objects spatial distribution & Virial distribution && Statistically subvirial \& identical \& supervirial compared to stars \\
        \ffc{} velocity distribution & Virial distribution && Statistically subvirial \& identical \& supervirial compared to stars \\
		Half-mass radius, $\halfmass$ && 1~pc & \\
		Virial radius && 1.3 pc & \\
        \hline
        \hline
        General characteristics && & \\
        \hline
		$N$-body (H\'enon) time unit, $T_*$ &  & 0.27~Myr & \\
		Crossing time, $t_{\rm cr}$ & & 0.25~Myr  &  \\
		Half-mass relaxation time, $t_{\rm rh}$ & & 40.63~Myr  & \\
        Stellar evolution \citep{kamlah2022a} & Mass loss enabled &&  \\ 
		External tidal force & Solar neighborhood &&  \\
		Simulation time && 1\,000~Myr &  \\
		\hline\hline
	\end{tabular}}
\end{table*}

We investigate the dynamical evolution of a populations of \ffc{}s in open clusters. We dynamically evolve three star cluster models (C025, C05, and C075), each containing $\nstar=11\,000$ stars, and $n_{\rm ffc} = 30\,000$ \ffc{}s. Among the stars, $n_{\rm bin} = 1000$ are in binary systems \citep[according to the number of binaries expected from these systems: see, e.g.,][and references therein]{offner2023}.
We adopt the method \cite{sana2012} for pairing the binaries. The semi-major axis of the binaries ranges from $0.01$ to 100~au.
Stellar masses are obtained from the \cite{2001MNRAS.322..231K} initial mass function (IMF) in the mass range $0.08-150\,\msun$. The total mass of each star cluster is $\clustermass\approx 6.62 \times 10^3 \msun$. The initial conditions of these models are listed in Table~\ref{table:details}. We generate our model in a modified version of Mcluster \citep{kupper,agostino}, which generate free-floating \fff{comets} along with the stellar population.

The positions and velocities for the stars are drawn from a \cite{Plummer:1911aa} model in virial equilibrium, with an initial virial radius of $\rvir=1.3$~pc, corresponding to an initial half-mass radius of $\halfmass=1.0$~pc. The stars share the same density distribution in all models (i.e., they are generated with the same random seed). The positions of the \ffc{}s are also drawn from a \cite{Plummer:1911aa} model, using the same characteristics except for the mass of the objects (and no IMF). \fff{We use a similar distribution as we expect the \ffc{}s to come from the protostellar phase of the stellar evolution, thus being a byproduct of stellar objects.} Moreover, the velocities of the comets are scaled to obtain the desired virial ratio \fff{( $Q = \ \mid T/U \mid$, where $T$ is the total kinetic energy and $U$ is the total potential energy of the system),} for our different models. 
At the generation of the \ffc{} density distribution, all comets are assigned a mass. The kinetic energy is changed then according to the wanted virial ratio of the \ffc{}s population. Like the stellar generation, the kinetic energy is differently distributed, even though we use the same random seed.
The \ffc{}s in model C025 are assigned a sub-virial distribution ($Q = 0.25$, where \ffc{}s have, on average, lower velocities than the stellar distribution), those in model C050 are assigned virial equilibrium ($Q= 0.5$), and those in model C075 are assigned a super-virial distribution ($Q=0.75$, where \ffc{}s have, on average, higher speeds than the stellar distribution). \fff{We adopt these initial conditions to determine how the initial velocity distribution of \ffc{}s affects their dynamical evolution over time and the mass segregation process.}
The stellar population is in virial equilibrium in all models.

Each star cluster evolves in an external tidal field corresponding to that of a Solar orbit in the Milky Way. The Milky Way is modeled as a point mass of $M_G=9.56\times 10^{10}\,\msun$, with the star cluster in a circular orbit with radius $R_G=8.5$~kpc. The corresponding tidal radius ($r_t$) of a cluster can then be estimated using $r_t\approx(\clustermass/3M_G)^{1/3}R_G$.

We use \nbm{}\footnote{\url{https://github.com/nbody6ppgpu/Nbody6PPGPU-beijing/tree/Nbody6PPGPU-ML}} \citep{flamminidotti2025} to carry out the simulations. As the \ffc{}s do not exert a gravitational force on the other bodies, they do not affect the evolution of the stellar population and the other \ffc{}s. Stellar evolution and binary evolution are implemented following the prescriptions of the stellar evolution package and their improvements \citep{Eggleton:1989aa, Hurley:2000aa, Hurley:2002ab, Hurley:2005aa, Belczynski:2007aa, kamlah2022a,spurzem2023}.

We integrate the star cluster models for 1\,000~Myr ($\approx \ 25 t_{\rm rh}$, see next section). Other important properties of the models, including the initial crossing time and initial half-mass relaxation time, are summarized in Table~\ref{table:details}. \fff{The \ffc{}s are treated as test particles during the simulations, unless noted otherwise.}


\section{Dynamical evolution of \ffc{} populations} \label{section:results3}

\subsection{Timescales}

The dynamical evolution of star clusters in this study remains predominantly influenced by the stars, as their mass is several orders of magnitude greater than that of the \ffc{}s. Thus, the initial crossing time, $\tcross$, and its initial half-mass relaxation time, $\trelax$ depend on the stellar properties. The crossing time for the model (since we use the same stellar distribution in all the three models) is 0.25~Myr (see Table~\ref{table:details}).  The half-mass two-body relaxation time \citep{Spitzer:1987aa} is defined as
\begin{equation}
    t_{\rm rh} =  \frac{0.138\nstar\halfmass^{3/2}}{\sqrt{G\clustermass}\ln\Lambda} 
    \quad .
\end{equation}\label{eq:relax}
Here, $\halfmass$ is the half-mass radius, $\ln\Lambda\approx \ln(0.02\nstar)$  is the Coulomb logarithm for different population masses \citep{Giersz:1994aa}, and $\nstar$ is the number of stars. In physical units, the initial half-mass relaxation time is 40.63~Myr. The mass segregation timescale in the generic model is $t_{\rm ms} = \trelax \ m_{\rm av}/m_{\rm max} \approx 0.006\trelax \approx 0.24$~Myr. For reference, the disruption timescale of the star cluster \citep{Lamers:2005aa} is approximately 6.6~Gyr.
In the following sections, we analyse an alternative approach to the mass segregation timescale, which will be used in order to analyse the long-term dynamics of the smaller size objects and to confirm \fff{whether}, with $N$-body direct simulations, is indeed possible to observe mass segregation of these small-mass objects.

\subsection{Lagrangian radii}\label{sec:lr}

\begin{figure}
    \begin{tabular}{c}
        \includegraphics[width=1.05\columnwidth]{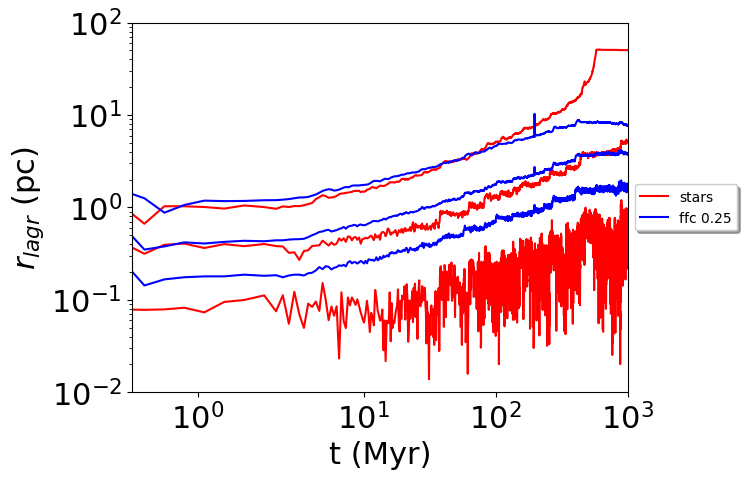}\\ 
        \includegraphics[width=1.05\columnwidth]{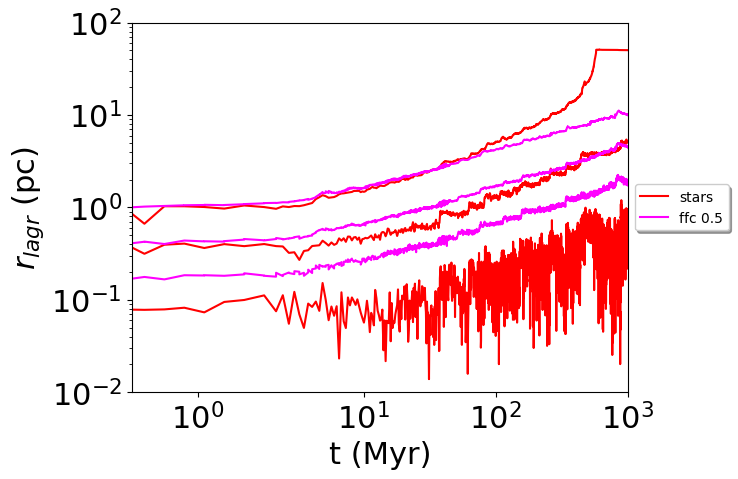}\\
        \includegraphics[width=1.05\columnwidth]{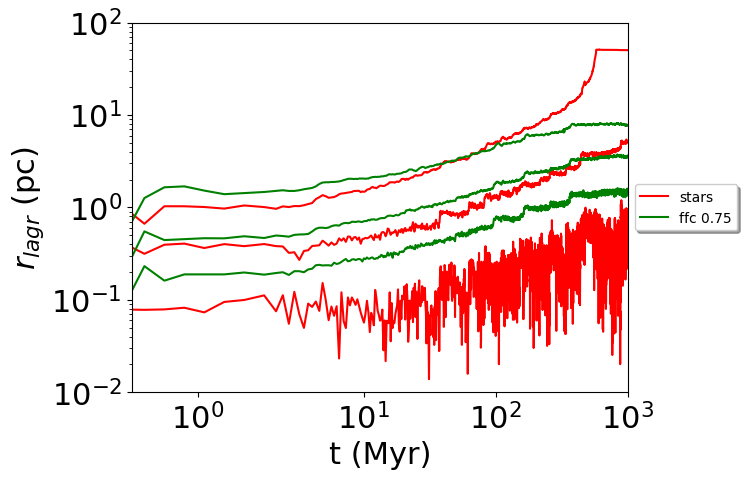} 
    \end{tabular}
    \caption{Evolution of the Lagrangian radii enclosing 0.1\%, 10\%, and 50\% of the total mass for stars and number percentage of \ffc{}s (indicated in the legend with the virial ratio of the model), for the models C025 (top), C05 (middle) and C075 (bottom). At each time step, the Lagrangian radii are calculated based on the total mass of the star cluster at that time. %
    }\label{fig:lagr0150}
\end{figure}

\begin{figure}
    \centering
    \begin{tabular}{c}
    \includegraphics[width=1.0\columnwidth]{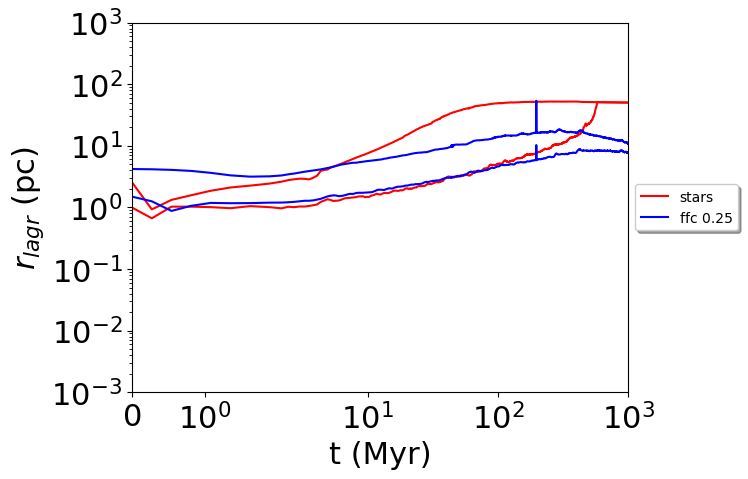} \\  
    \includegraphics[width=1.0\columnwidth]{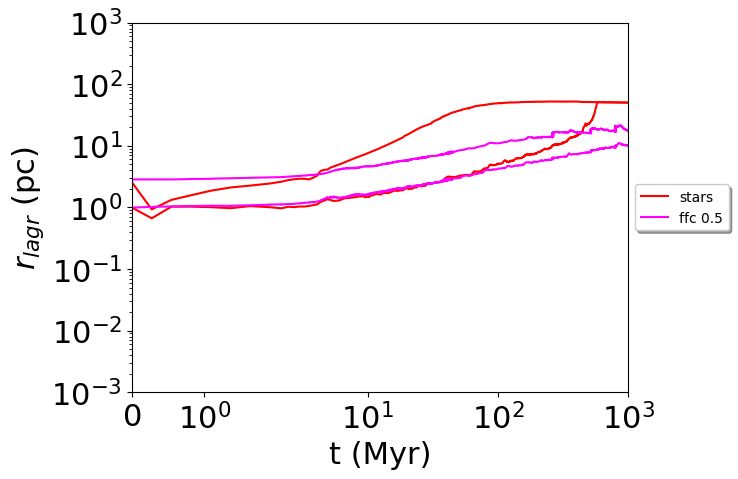}\\  
    \includegraphics[width=1.0\columnwidth]{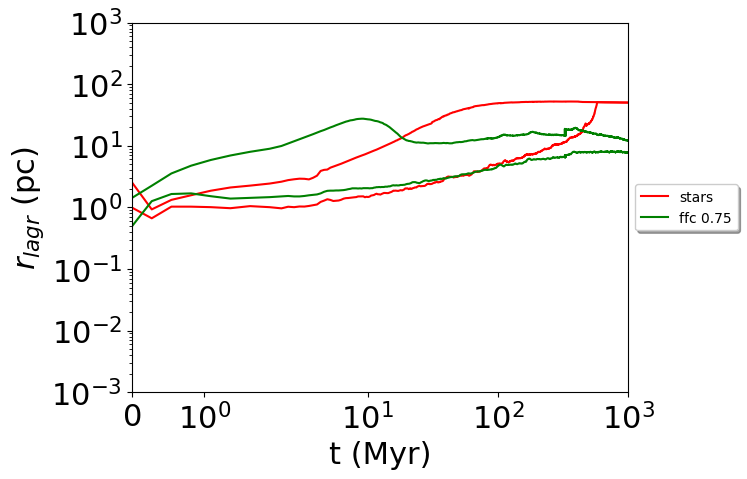}
    \end{tabular}
    \caption{Same as in Figure~\ref{fig:lagr0150}, for the 50\% and 90\% Lagrangian radii. 
    } \label{fig:lagr5090}
\end{figure}

\begin{figure}
    \centering
        \includegraphics[width=1.0\columnwidth]{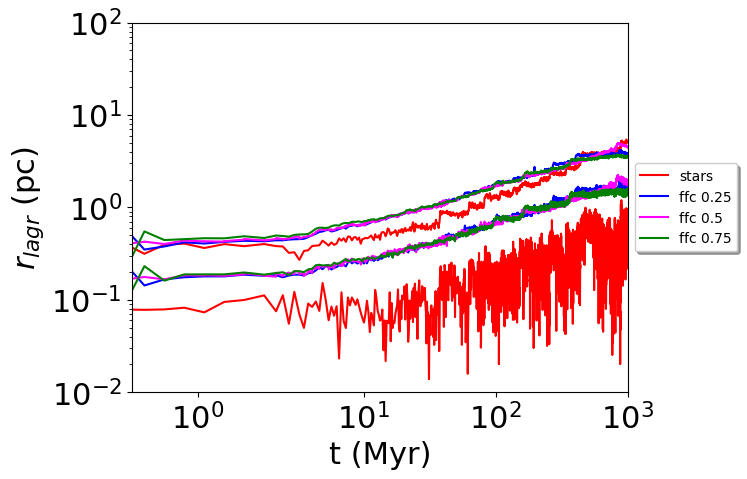} \\  
    \includegraphics[width=1.0\columnwidth]{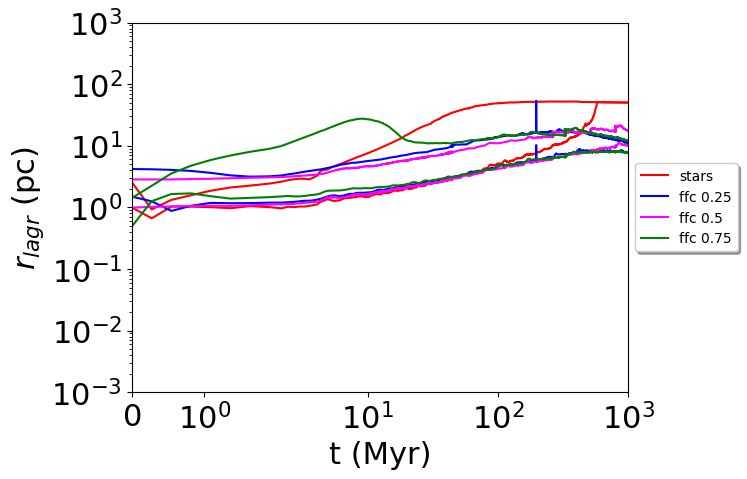} \\  
    \caption{ {\it Top:} evolution of the Lagrangian radii containing  the percentage of mass of 0.1\% and 10\% for both stars and number percentage of \ffc{}s from each model. {\it Bottom:} same, but for the 50\% and 90\% Lagrangian radii. The Lagrangian radii at each time are calculated using the \fff{total mass of the star cluster at that time}. } \label{fig:allmodelslagr}
\end{figure}

Lagrangian radii provide a powerful tool for analyzing the global evolution of a star cluster. The evolution of the Lagrangian radii for all simulated models is depicted in Figures~\ref{fig:lagr0150}, \ref{fig:lagr5090}, and~\ref{fig:allmodelslagr}. 
For the peaks in the \ffc{}s Lagrangian radii, they are due to an \ffc{} particle that is \fff{on} the verge of escaping the cluster. For a more detailed explanation, please check the previous paper of this series \citep{flamminidotti2025}.
At each moment in time, the Lagrangian radii for the stars are calculated relative to the total cluster mass at that time. 
The Lagrangian radii for the massless particles depend solely on the number of \ffc{}s within the Lagrangian shell at the actual time, as all \ffc{}s have the same mass.

Over time, the star cluster undergoes gradual expansion and mass loss, driven by stellar evolution and the escape of stars from the cluster. It is important to note that the stellar Lagrangian radii, as shown in Figures~\ref{fig:lagr0150} and~\ref{fig:lagr5090}, are independent of the properties of the \ffc{} population. Since \ffc{}s are massless, their presence does not influence the dynamical evolution of the stellar population.

For the stellar population, at $t \approx 8$~Myr, the cluster core undergoes a slight contraction, followed by a rebound. This event triggers strong scattering, resulting in the ejection of stars to the outskirts of the star cluster and beyond. Simultaneously, stellar evolution contributes to a gradual reduction in the cluster's total mass. Over the course of approximately 20~Myr, these processes lead to the loss of about 10\% of the cluster's initial total mass. As a result of stellar ejections and stellar mass loss, the gravitational potential of the star cluster decreases. \\

The dynamical evolution of the \ffc{} population is linked to the evolution of the core of the star cluster \citep{flamminidotti2025}. In the inner regions of all models, the \ffc{}s exhibit a similar orbital behavior around the core (Figure~\ref{fig:lagr0150}).
During the first 0.5~Myr, the innermost regions of the \ffc{}s population undergo contraction and expansion in models C025 and C075, respectively. \fff{This is driven by the contraction due to the low average  velocity of the \ffc{}s and the escape of high-velocity \ffc{}s in models C025 and C075, respectively}. 
The 50 \% shells of the \ffc{}s population behave similarly, in all models, after $\sim$ 1~Myr. After this initial phase of contraction and expansion in the two different models, the inner regions behave mostly similar for all models. 
In the 0.1 \% shell, the \ffc{}s tend to orbit the core at a larger distance than the stars. We have a similar evolution also for the 50 \% shell, which is slowly expanding compared to the stellar 50 \% shell.

In the outer regions (Figure~\ref{fig:lagr5090}), the behavior is similar for all the three models, but we experience a retraction in both C025 and C075 models. Since \ffc{}s have already lost more than 50 \% of particles and possibly the highest energy members have been ejected, the 90 \% shell of \ffc{} is contracting rather than expanding. 

However, \fff{this shell clearly tends to shrink around 800~Myr in models C025 and C075, while the model C050 (initially in virial equilibrium) also begins to contract, although less noticeably (extending the simulations by another 20~Myr confirms a similar trend). This behaviour arises from the gradual loss of kinetic energy among the \ffc{}s and the occurrence of ejections. Since model C075 experiences slightly more ejections (see Section~\ref{sec:escapers}), the outer \ffc{}s are more strongly affected by the gravitational potential of the core and move inward. Similarly, in model C025, the total kinetic energy of \ffc{}s is lower than in the other two models, so they are more efficiently pulled toward the centre. At this stage, the average velocity of the \ffc{}s is low enough that they become influenced by the deeper potential well of the core, causing them to move inward and remain confined in the inner regions.}

\begin{figure}
\begin{tabular}{c}
  \includegraphics[width=\columnwidth]{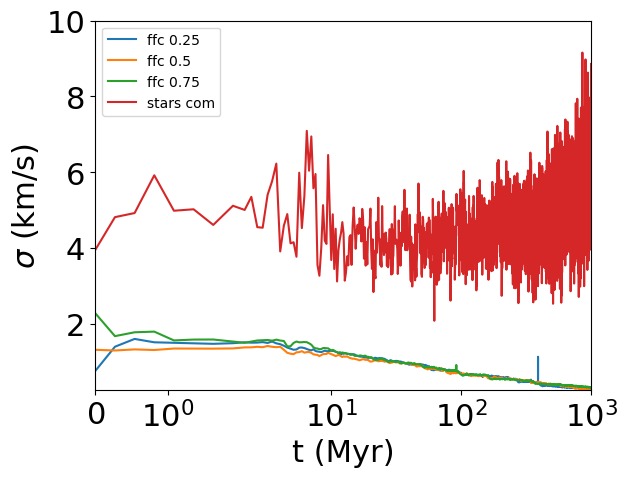}
  \end{tabular}
  \caption{Velocity dispersion of the stellar population and the \ffc{}s population. The \ffc{}s  velocity dispersion is smaller than the stellar population in all models. \ffff{The presented velocity dispersion consider the center of mass for the binary systems.} A zoom in on the \ffc{}s curves is in Appendix \ref{app4}.
  }\label{fig:disp}
\end{figure}
\fff{The velocity dispersion shown in Figure~\ref{fig:disp} confirms our previous conclusions. The stellar population has a larger velocity dispersion compared to that of the \ffc{} population, and the velocity dispersion of the \ffc{}s gradually reduces in all models. The initial imprint of the different velocity distribution is clearly present in models C025 and C075, but it rapidly vanishes within less than a Myr.} 
The results also imply that \ffc{}s do not follow mass segregation.\\
One may wonder whether the \ffc{}s would participate in the energy equipartition, since there is a large difference in energy between the population, or similar energy, respectively in the C025 and C075 models. 
\fff{The mass difference between the components in a star cluster is significant when also considering planets or comets. If we only include stars (e.g., $0.08~\msun$ to $\approx 60~\msun$), the mass ratio is roughly two orders of magnitude. Compared to the average mass ($\approx 0.58\ \msun$), the mass ratio is typically one order of magnitude. 
Planets and comets, on the other hand, are much less massive than stars. Such bodies are dynamically irrelevant for the evolution of the star cluster.
Moreover, with the global dynamical evolution of \ffc{}s, we have shown that this phenomenon is not energy dependent, as all three models have a relatively similar behavior after just several Myr. }


\subsection{Kinematical evolution of \ffc{}s in star clusters}\label{sec:kffcs}

\begin{figure}
    \begin{tabular}{c}
    \includegraphics[width=\columnwidth]{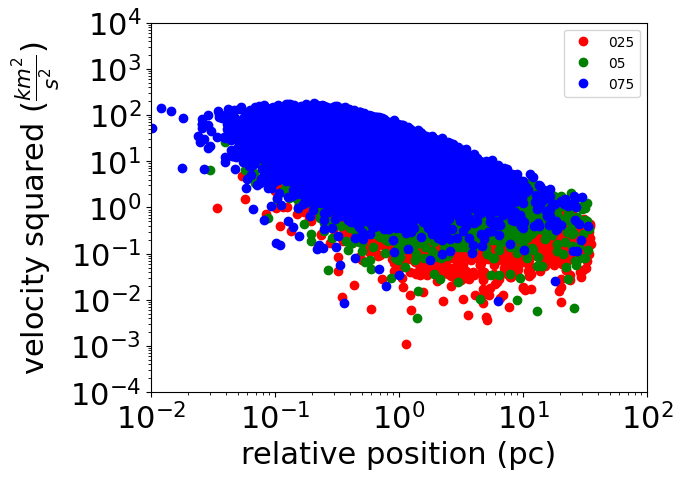}\\
    \includegraphics[width=\columnwidth]{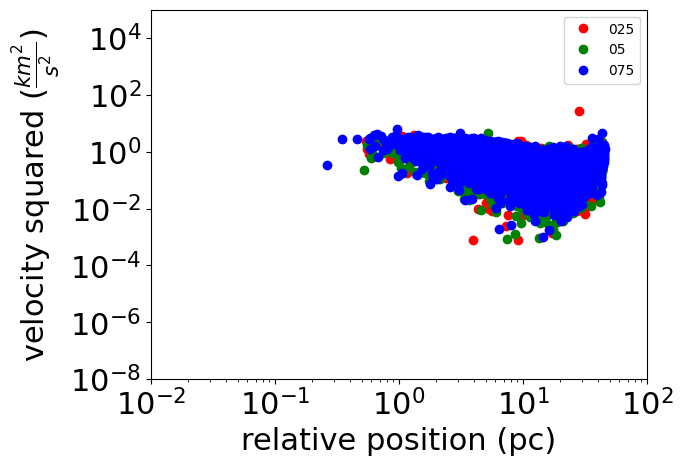}
    \end{tabular}
    \caption{Radial positions and square velocities of \ffc{}s, for all simulated models, at $t=0$~Myr ({\it top panel}) and at at the end of the simulation at $t=1\,000$~Myr ({\it bottom panel}). \fff{We also added every single model separately from the rest in Appendix \ref{app3}.}
    }\label{fig:posvsvel2}
\end{figure}

We simulated models in which the \ffc{}s have different initial \fff{velocities}; therefore the distribution resulting for equal-mass \ffc{}s depends mainly on their speeds. The speeds differ between the three models, as shown in Figure~\ref{fig:posvsvel2}. The square velocity of these system at the start of the simulation (top panel) is quite different, with  speed ranging between the maximum velocity of model C075 ($\approx 13.36 $~\kms) and the minimum speed of model C025 ($\approx 0.03$~\kms). The average initial speeds are 1.91, 3.31 and 5.73 \kms for model C025, C05 and C075, respectively. After 1\,000~Myr (bottom panel), the \ffc{} distributions of all models are similar. The same plot, with separated \ffc{} for all models, is present in Appendix~\ref{app2}.
The minimum speed of models range from a minimum of $2.79 \times 10^{-2}$ \kms for model C025 to a maximum of 5.19~\kms{} for model C025. The average values are 0.76, 0.76 and 0.85~\kms{} for model C025, C05 and C075, respectively. \fff{Appendix~\ref{app2} shows the distributions for each model.}

Three key conclusions can be drawn from \fff{comparing the \ffc{} distributions at the start and at the end of the simulations:} (i) the distribution of particles is quite similar for all models at the end of the simulation, which tell us that the imprint of the initial conditions disappeared, (ii) the position of the remaining particles is mostly moving in the inner orbits, while the outer regions have barely changed, testifying that the core is pulling the \ffc{}s in. The mass of the core remains quite similar for all the models (\fff{see Appendix~\ref{app1}} the radius and mass of the core in the star cluster for model C05), and (iii) the high-velocity \fff{\ffc{}s have escaped from the cluster, leaving those with speeds below roughly  1~\kms. The latter observation is especially important, as we can assume that any \ffc{}s with sufficient initial speed} to escape from the inner regions of the cluster, ultimately did. For our initial conditions, the escape velocity from the cluster center at $t=0$~Myr is $v_{\rm esc} = 3.31$~\kms{}. Thus, close encounters with stars will eventually result in the ejection of objects at higher velocities within just \fff{several million years. Subsequently, other \ffc{}s are} ejected either from tidal stripping in the outer regions or evaporation. 


\subsection{Escaping stars and \ffc{}s}\label{sec:escapers}

\begin{figure}
\begin{tabular}{c}
  \includegraphics[width=\columnwidth]{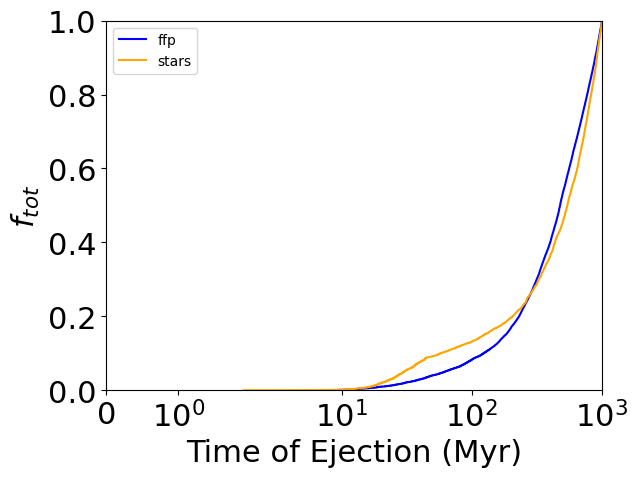}\\
  \includegraphics[width=\columnwidth]{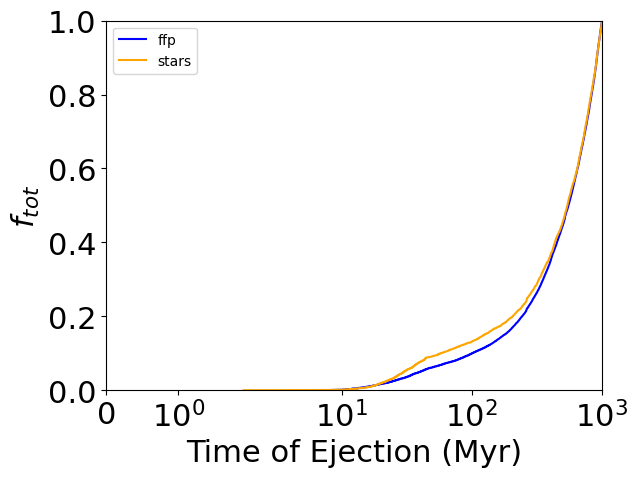}\\
  \includegraphics[width=\columnwidth]{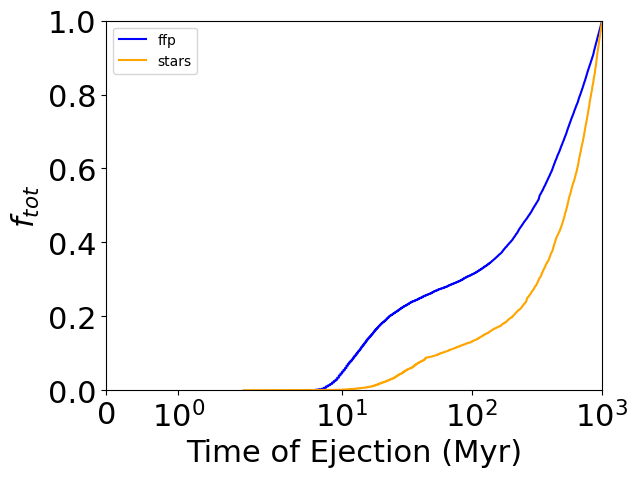}
\end{tabular}
  \caption{Cumulative ejection distribution of the escapers, including both stars and \ffc{}s. Notably, the ejection function for all three \ffc{} models shows variation only at the beginning, after which it progresses almost linearly.
  }\label{fig:escall}
\end{figure}

\begin{figure}
\begin{tabular}{c}
  \includegraphics[width=\columnwidth]{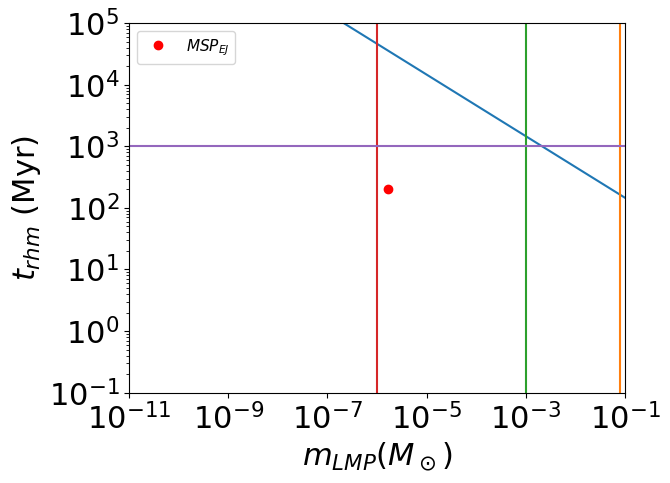}
  \end{tabular}
  \caption{Relaxation timescale of the LMP population for different masses, using our initial conditions for the number of LMP and the half-mass radius. The relaxation timescale for a Earth mass and an Jupiter mass are from the red line and green line respectively. The orange line is the star limit (i.e., the ignition of hydrogen at $\approx 0.08 \msun$). The red dot is considering a mass spectra of LMP in a uniform logarithmic distribution, ranging from Earth mass to Jupiter mass, using the same initial condition of our main models. Finally, the blue line is using the initial conditions of the model C05.
  }\label{fig:mvstrhm}
\end{figure}

\begin{figure*}
\begin{tabular}{c}
    \includegraphics[width=1\columnwidth]{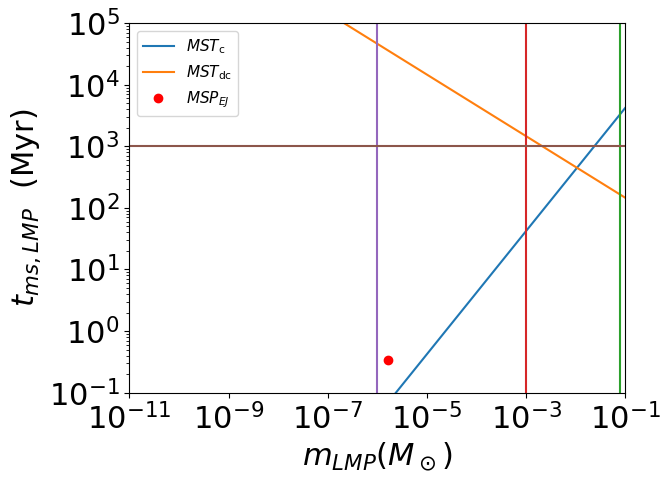}
        \includegraphics[width=1\columnwidth]{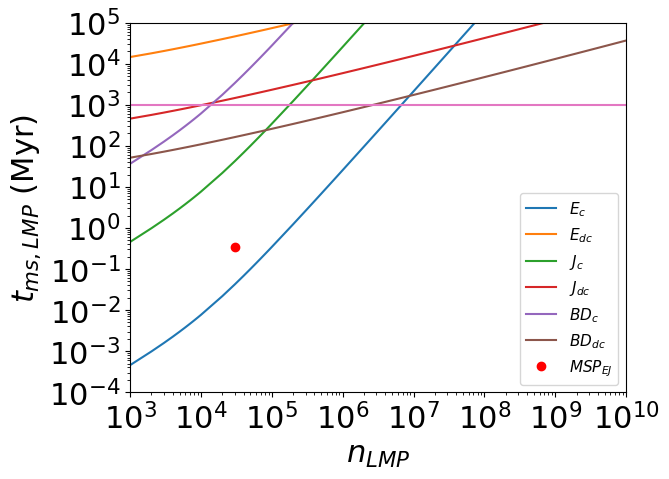}
  \end{tabular}
    \caption{Mass segregation timescale of the LMP population for different masses (left) and number of initial objects (right), using our initial conditions for the number of LMP and the half-mass radius. 
    In the left plot we find the mass segregation timescale for coupled ($MST_{\rm  c}$) and decouple ($MST_{\rm dc}$) timescales. $MSP_{\rm EJ}$ is instead the mass segregation timescale for a decoupled mass spectrum from Earth to Jupiter masses, distributed in a logarithmic uniform distribution.
    On the right, with similar semantics, we have Earth, Jupiter and Brown Dwarf coupled (c suffix) and decoupled (dc suffix).
    The relaxation timescale for a Jupiter mass and an Earth mass are indicated with the orange and green lines, respectively.
    }\label{fig:mvstms}
\end{figure*}

Figure~\ref{fig:escall} shows the cumulative fraction of ejected particles. The stellar ejection rates are identical for all three models. The total number of ejected stars at $t=1\,000$~Myr is 4\,736 (47.36 \% of the initial number of stars), while the total number of ejected \ffc{}s is 18\,355, 17\,625 and 18\,367, respectively, for models C025, C05 and C075 (corresponding to 61.18 \%, 58.75 \% and 61.22 \% of the initial total number of \ffc{}s).
The higher number of ejected particles is due to the relative low density of the star cluster in the initial conditions. 
Moreover, we can see that the total number of ejected \ffc{}s is  similar in all models, further justifying their independence from their initial conditions.
At the start of the simulation we can clearly see a major difference in ejection due to the energy distribution of the \ffc{}s, where the C075 model ejects 25 \% of particles in few million years, and then proceeds almost linearly until the end of the simulation. Model C025, instead, evolves as a logarithmic curve up until 300~Myr, and then proceed linearly. The model in virial equilibrium C05, instead, behave similarly to the stellar cumulative curve, with an initial difference due to stellar evolution.
Thus, we can conclude that the imprint of initial conditions is mainly observed at the start of the simulation, in a timescale less than a relaxation time.

Consistent with previous studies and with \cite{flamminidotti2025}, the ejection speeds of stars and \ffc{}s display the same larger velocities for stars and smaller for \ffc{}s during the ejection. As observed in earlier work of this series, the stars with the highest speeds are neutron stars (see Appendix \ref{app3}).\\

The average velocities of the ejected stellar population are 10.02~\kms{} for all models, while the corresponding values for the ejected \ffc{}s are 1.22~\kms{}, 1.29~\kms{} and 1.91~\kms{} for models C025, C05 and C075 respectively.
The slight difference in average speed is due to the initial ejection of \ffc{}s, which is slightly faster in model C075 and slightly smaller in model C025 and C05. This is a consequence of the diverse initial energy distribution, which gave larger velocities in model C075 and smaller velocities in model C025. Except for random fluctuations in the ejection velocities,  we have the same velocity distribution in the ejected particles.\\
\fff{In the following section, we present an alternative approach to the mass segregation timescale, which we will use in order to study the long-term dynamics of the lower-mass objects, and to determine with $N$-body simulations whether such populations indeed experience mass segregation.}


\section{Mass segregation with two different populations} \label{section:result4}

\subsection{Relaxation and mass segregation times for LMP}\label{sec:rlxlmp}

We saw that the \ffc{}s, in a long term dynamical evolution scenario, are either ejected through evaporation or for ejection at high velocities, \fff{mostly} due to the type of models we chose. \fff{In this section we take a more general approach, using the initial conditions of our previous models as an example, but not using the simulation data.}
Thus, the role of segregation of low mass particles (like comets) in a star cluster is an interesting topic. These particles do not follow mass segregation \citep[like shown in the previous paper of this series, see][ and here]{flamminidotti2025}. Thus, these particles should independently not segregate and be mostly governed by the inner regions, despite the number of particles and their mass.\\
Following Section \ref{sec:msdisc}, we will consider low mass particles anything that is not considered a star. We therefore consider, for the low mass particles, a maximum mass of 0.08 $\msun$ (brown dwarfs) and a minimum mass of the comets. We use the same terms LMP and HMP, since we will effectively describe several type of objects for LMP. \fff{We will use these terms until the end of Section~\ref{section:result4}}.\\
For all purposes, we use Equation~\ref{eq:relax} for the "classical" relaxation time. \fff{We use the following expressions for the mass segregation timescale:}
\begin{equation}
    t_{\rm ms,LMP1} = k N_{\rm LMP} \ln(0.02(N_{\rm LMP}+N_{\rm HMP})) \Delta m \ m_{\rm max}^{-1} r_{\rm hm}^{3/2} \quad,
    \label{eqtms1}
\end{equation}
\begin{equation}
    t_{\rm ms,LMP2} = k N_{\rm LMP}^{3/2} m_{\rm LMP}^{-1/2} N_{\rm LMP \& HMP}^{-1}r_{\rm hm}^{3/2} \ln(0.11(N_{\rm LMP})) \quad,
    \label{eq:tms2}
\end{equation}

\fff{which are the coupled and uncoupled mass segregation timescales, respectively. The derivation and interpretation of these expressions is provided in Appendix~\ref{sec:msdisc}.}

Figure~\ref{fig:mvstrhm} shows the evolution of the relaxation timescale in terms of the decoupled relaxation time, using the initial conditions of our reference models, excluding the LMP mass. 
For the LMP, using the same initial conditions of this work, we find that $t_{\rm rlx,LMP} \approx 3.46 \times 10^8$~Myr, which is four orders of magnitude larger than the Hubble time. 

A similar conclusion is found for terrestrial-like planets. Interestingly, using a mass spectra of a LMP population ranging from $1\,M_{\rm E}$ to $1\,M_{\rm J}$ in a logarithmic uniform distribution, the timescale for relaxation is quite low for these masses. 

Finally, for our model, we see that the LMP have an acceptable timescale (on the order of a Gyr) for masses near the stellar limit. This approximation of the relaxation timescale is valid only for low masses, as otherwise the stellar population would not decouple from the LMP. 

The previous results give us two important conclusions, (i) the mass spectra for a planetary population, like shown in the figure, may effectively be important to look over for how the LMP evolves dynamically and (ii) near stellar masses, such as brown dwarfs, are likely to follow mass segregation in comparable timescales to that of stellar populations. The former point is interesting in its own "internal segregation", in which we may check how the Earth-sized planets evolve as compared to the Jupiter-like planets.

We therefore wish to determine the threshold mass of a LMP to be not decoupled from the stellar population. We can analyse this limit in terms of mass segregation timescale in the next section.
In there, we will consider the relaxation time of the decoupled population, but we consider the average stellar mass of the cluster used in our model.
If we will consider the decoupled mass segregation timescale, than the term $m_{\rm av}/m_{\rm max}$, with same-mass LMP, will be unity. 

\subsection{Analysis of the decoupled mass segregation timescale}\label{sec:MSTD}

From the timescales we discussed in \fff{Appendix \ref{sec:msdisc}}, we can effectively compare and analyse the mass threshold value for the particles to be included in the mass segregation process. 
There is a strong dependence, in the LMP, with the number of their objects and their mass. Smaller $N_{\rm ffc}$ and larger $m_{\rm ffc}$ would result in a smaller timescale. Although not analysed in this work, the (initial) spatial and velocity distribution of LMP is also rather important, especially if the LMP are formed in a, overall, smaller region of the cluster.

Our results are similar to same-mass clusters, where mass segregation is larger cause of the same mass-elements. However, compared to that case, there is no stellar evolution for the LMP. \\
In Figure~\ref{fig:mvstms} we show the mass segregation timescale as a function of the mass of the LMP and the number of LMP respectively in the left and right plot, assuming the star cluster initial condition used in this work. We immediately can tell that the coupled mass segregation term does not bring result according to any result we obtained so far in the literature, and that mass segregation timescale does not agree with our simulations (and it partly do not agree with brown dwarfs too, as we will see in the next section). The decoupled mass segregation seems to be more in agreement with our results on the relaxation timescale. 
To obtain a broader understanding of the possible scenario, we list the result observed for the different LMP used in Figure~\ref{fig:mvstms}. 

\begin{itemize}
    \item If we use comet-mass objects, their mass can be any dimension smaller than a planetary mass. A mass average value would be $10^{-14}\,\msun$, while we constructed more massive comets, on the order of $10^{-11}\,\msun$ for this simulation. Within our initial setting of initial conditions, the mass segregation coupled timescale is too large ($t_{\rm ms,LMP} = 4.61 \times 10^8$\,Myr). The same is true for larger number of LMPs, as the number grows exponentially (linearly on a log scale).
    \item If we use planet-mass objects, we have $t_{\rm ms,LMP} = 1.46 \times 10^3$~Myr and $4.61 \times 10^4$~Myr for a Jupiter-like mass and Earth-like mass, respectively. Earth-like objects shares the same issue as comets, with a timescale larger than an Hubble time. The Jupiter mass planets seems the most suitable candidates for our initial conditions, instead. 
    \item For brown dwarfs, the mass segregation timescale is $t_{\rm ms,LMP} = 1.63 \times 10^2$~Myr, being thus the more accurate candidate to test the mass segregation limit (see next section).   
    \item The case of mass spectra is particularly interesting, and worthy of testing. \fff{Whether} the masses are gravitationally pulled by the inner regions similarly to other cases or they manage to relax before this occurs, is the main idea for the next work.
\end{itemize}

In the previous list of results  we opted out the HMP mass segregation timescale, but this is indeed an important factor. Since these population will segregate faster, \fff{we expect the cluster to have significantly evolved away from its initial conditions after the cluster relaxed.} This may be especially useful in low density star clusters. We will consider this in future works.
Thus, the LMP with masses significantly smaller than those of stars become decoupled and exhibit very long mass segregation timescales.
In our previous assumptions, we left the half-mass radius as similar in both cases. A more concentrated distribution in the center or a more enlarged distribution, as seen in  Sections~\ref{sec:lr} and~\ref{sec:kffcs}, would just remove particles in the outermost regions, while objects would continue to orbit in the cluster center. The assumption of similar initial conditions between \ffc{}s and stars makes more sense than a completely diverse one, as comets are a byproduct of stellar and proto-planetary disks. The initial energy, as we saw in the previous section, do not diverge too far away from the final results. Thus, a final confirmation is needed, in terms of a mass spectrum for the planets. 

Here, we also carried out a simulation using brown dwarfs instead of comets, as we are going to see in the next section. We plan to analyse the upper limits of "non-stellar" objects to observe the dynamical behaviour of the latter for near-stellar mass objects. %


\subsection{Dynamical evolution of brown dwarfs and asymptotic mass segregation}\label{BDAMS}

\begin{figure}
\begin{tabular}{c}
  \includegraphics[width=\columnwidth]{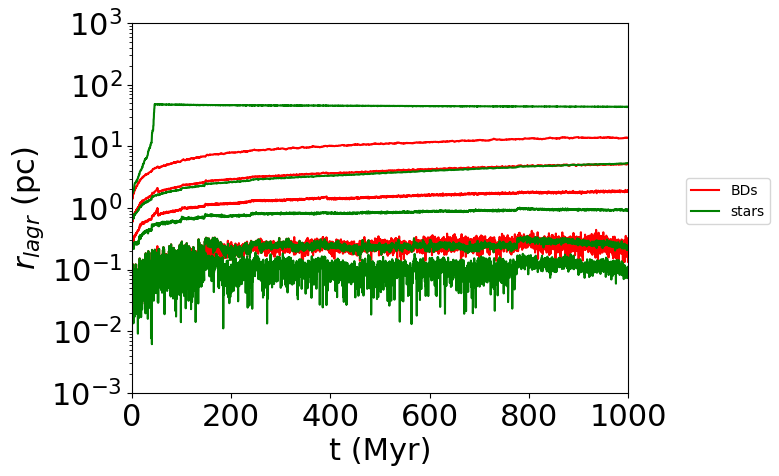}\\
  \includegraphics[width=\columnwidth]{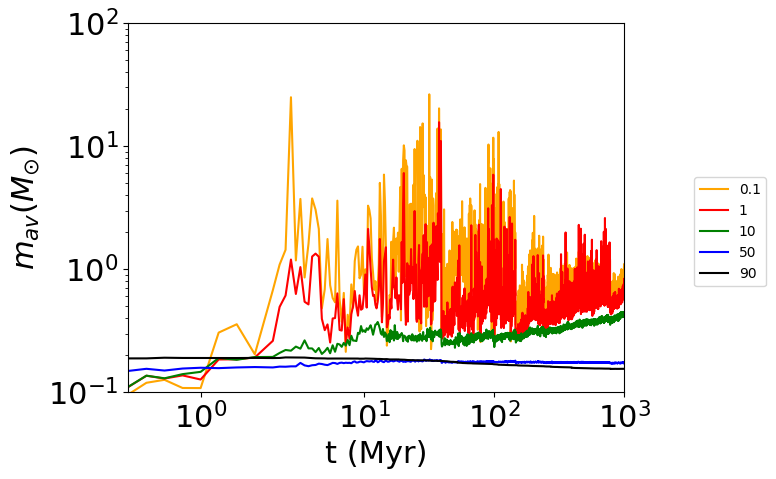}
\end{tabular}
  \caption{In the top figure, the Lagrangian radii evolution of both stellar and brown dwarf components in the 0.1 \%, 1 \% ,10 \%, 50 \% and 90 \% shells. The lagrangian radii evolution is green for the stellar components and red for the brown dwarfs. In the bottom figure, with identical shells, we display the average mass evolution.
  }\label{fig:bdevol}
\end{figure}

The results coming from a simulation of brown dwarfs, using the initial conditions of Table \ref{table:details} but the mass of the \ffc{}s, gives extremely interesting results.
For a more direct comparison of what was argued in the previous section, we show the dynamical evolution of brown dwarfs in Figure~\ref{fig:bdevol}, using the same set of initial conditions of Table~\ref{table:details}, except for the mass of the "massless objects". 
We used the classic version of \nb{} for this simulation. We do so because brown dwarfs are extremely near the stellar mass, and thus the massless approximation we previously apply is not eligible anymore.\\
The number of ejected stars and brown dwarfs is lower than the stellar and \ffc{} counterpart, as the model is extremely more massive overall. The 14.48\% of stars is ejected and the 16.18 \% of brown dwarfs is ejected at the end of the simulation. \\
We observe a similar evolution of the Lagrangian radii compared to that of the comets. \fff{However, t}here are \fff{also }several differences: (i) around 80~Myr, we see a peak in the outer regions, which suggest a partial mass segregation of these objects (like suggested in the bottom plot of  Figure~\ref{fig:bdevol}),  and (ii) the populations in the inner regions are more mixed, and the bump in average mass around 200 Myrs is in line with the mass segregation decoupled timescale. It is quite interesting to see that indeed there is a second inverse segregation, that is a direct consequence of the mixing between the two population. %
This is due to the total mass of the brown dwarfs being comparable with the core mass. In this specific case, it is also slightly larger by a factor of 100 $\msun$. Since the masses are comparable, there is a slight mixing, with some brown dwarfs orbiting around the core, and some segregating in the outer regions. Although not extremely precise, the brown dwarfs are indeed the mass limit for an object to effectively segregate in a cluster. A different set of initial conditions (for example a larger number of brown dwarf\fff{s}) would make this difference more important.\\
Thus, our previous approach on the mass segregation timescale is only partly correct, as it does not take into account the mass difference in the populations, though these are treated as decoupled. This is related to a thermodynamical behaviour of the system.\\
\fff{We now consider again a general approach, using HMP and LMP for the cluster populations.}
According to \cite{gomezleyton2019}, the mass segregation of a multi-populated mass system such as ours, the mass segregation of the smallest component does not behave as in the classical way, but asymptotically. If the $m_{\rm core,HMP} \gg N_{\rm LMP} m_{\rm LMP}$, where $m_{\rm core}$ is the core mass of the HMP, then the secondary population should not segregate. Thus, mass segregation effectively occurs when the total mass of the LMPs is closer to that of the inner regions.
Thus, we expect the system to not segregate even a larger masses. 
This conclusion affects the astrophysical implications of this study, and it justify the results shown in Section 3. 
Thus the core mass is the effective thermodynamical catalyst for the evolution of our LMP. This cannot happen for planet-size objects, and it would only happen for near stellar-size components (similarly to multiple stellar populations in globular clusters).
Thus, both observed open clusters and globular clusters are not ideal candidates to have mass segregated populations:
\begin{itemize}
    \item Open clusters are less dense, but have extremely short timescales for cluster dissolution. A near dissolution star cluster candidate may be the best case to observe whatever the LMP would segregate in that case.
    \item Globular clusters have extremely long dissolution timescales, and thus the cluster would remain much more massive than the LMP population.
    \item Specific cases, such as sub-structured clusters or tidal tails, would be interesting prospects to observe this population as segregated, \fff{at least}, locally.
\end{itemize}
The only remaining factor to understand is whatever a mass spectrum of planets is a redundant idea or not. The same work of \cite{gomezleyton2019} have a dependence on both the total mass and the single mass of the components of the two populations. Thus, more un-equal populations may direct this study in another direction and prove if these systems are both decoupled and are only dependent by the core mass, even with a different mass distribution. An internal segregation, of only the planet-sized particles, could bring a different aspect on this topic. We will focus on the latter for the next work of this series. 
%




\section{Conclusions} \label{section:conclusions}

In the first paper of this series \citep{flamminidotti2025}, we explored the dynamical evolution of massless particles. We concluded that, even with varying cluster densities, the dynamical evolution is primarily determined by the core's evolution rather than being influenced by mass segregation. In this second paper, we instead modify the energy distribution of the massless particles, which, on a technical level, corresponds to varying their kinetic energies. Moreover, we analyse of the kinematics of \ffc{}s and their escape rates. Finally, we show the mass segregation limits for massless particles as compared to the respective stellar component, highlighting the thermodynamical aspect of the system\fff{, resulting in the HMP core being the} main catalyst for \fff{the} mass segregation \fff{of the LMP}. Given our conclusion, there are only extremely specific cases in which we could observe segregation of free-floating comets or planets. The main results of our study can be summarised as follows:
\begin{enumerate}
    \item The spatial distributions and kinematics of the three models of \ffc{} population diverges at the start for the non-virialised models, which virialise shortly after. This effect is quite fast, on the order of 0.5~Myr, which is $\approx 2 t_{\rm ms}$. This process is mainly related to the high velocity \ffc s being ejected in model C075 and to the stellar encountering with other stars in model C025. These events, which substantially change the overall kinetic energy of the particles. The system becomes virialised in a short timescale, which means that mass segregation is not achieved in faster (or shorter) times for different energies.
    \item Interestingly, the 90\% Lagrangian shell of model C025 and C075 both stars to contract around 25 relaxation times. The C05 model also contract some Myrs later. This is related to the particles being attracted to the core, when high energy \ffc s are ejected and a large part of the \ffc s are evaporated, leaving just low velocity objects. Moreover, the \ffc{} population drastically dropped to $\approx$ 40\%, resulting in a less populated, and thus less diversified, \ffc{} population, which lost any imprint of the initial energy distribution.
    \item Brown dwarfs follows a different path to comets, as their results hints a more mixed dynamical behaviour for these objects, orbiting around the core like the \ffc{}s and be also mixed in the innermost regions like stellar components. An interesting result is a inverse mass segregation around 200 Myr, where most brown dwarfs are mixed in the innermost regions.
    \item The study on the mass segregation of massless particles follows and verifies the conclusions of \cite{gomezleyton2019}, and prove that the stellar mass in the star cluster core is fundamental for the evolution of the massless particles. The system relaxes violently at the start for non-virialised systems, but this is related mostly due to the initial conditions (i.e., velocities) of the \ffc s.
    \item Spatial separation ("mass segregation") of \ffc{}s is highly unlikely in open clusters (which dissolve quite rapidly) but interesting to check in dissolution-like scenarios. Globular clusters (which dissolve too slowly and have core masses significantly higher than the population of \ffc{}s) are improbable scenario.
\end{enumerate}

We have not included any planetary companions among star cluster members. Realistic planetary systems may alter the orbits of captured \ffc s through gravitational interaction during periastron passage. Although integrating multi-planet systems in a comparatively short time is still challenging in N-body direct simulation, substantial progress has been made in the recent decade \citep[e.g.,][]{Malmberg:2007aa, Hao:2013aa, Shara:2016aa}, and combining simulations of star clusters with those of planetary systems is now possible using the \amuse{} software environment \citep[see][]{Portegies-Zwart:2013aa, Pelupessy:2013aa, Cai:2016aa} through sequentially integrating the 
stellar and planetary components
\citep[e.g.,][]{Cai:2015aa, Cai:2017aa, dotti_cai_spurzem_kouwenhoven_2018, 2019MNRAS.489.2280F, flamminiblack,stock2020,flamminidotti2023,wukai2023,wukai2024}. 
This is a necessary step for obtaining a full understanding of \ffc{} populations in star clusters, in particular for understanding the transfer of material between planetary systems through physical collisions between planets and exo-comets.

Finally, large numerical simulations with a large number of \ffc{}s, would solve this paradigm definitely. However, there are still not good enough direct $N$-body codes (or super-computing resources) to reach the minimum possible ($\approx 10^{8}$ \ffc{}s) in order to reach a plausible number of \ffc{}s for these simulations.


\begin{acknowledgements}

\fff{We thank the anonymous referee for the comments, which helped to improve dramatically the quality of our work.}
This material is based upon work supported by Tamkeen under the NYU Abu Dhabi Research Institute grant CASS.
FFD, RS, and KW acknowledge support by the German Science Foundation (DFG, project Sp 345/24-1).

AA acknowledges support for this paper from project No. 2021/43/P/ST9/03167 co-funded by the Polish National Science Center (NCN) and the European Union Framework Programme for Research and Innovation Horizon 2020 under the Marie Skłodowska-Curie grant agreement No. 945339. This research was funded in part by NCN grant number 2024/55/D/ST9/02585. For the purpose of Open Access, the authors have applied for a CC-BY public copyright license to any author Accepted Manuscript (AAM) version arising from this submission.
PB thanks the support from the special program of the Polish Academy of Sciences and the U.S. National Academy of Sciences under the Long-term program to support Ukrainian research teams grant No.~PAN.BFB.S.BWZ.329.022.2023.
MG was supported by the Polish National Science Center (NCN) through the grant 2021/41/B/ST9/01191.
KW and RS acknowledge support by German Science Foundation (DFG) under grant Sp 345/24-1. KW and RS acknowledges NAOC International Cooperation Office for its support in 2023, 2024, and 2025, and the support by the National Science Foundation of China (NSFC) under grant No. 12473017. 
This research was supported in part by grant NSF PHY-2309135 to the Kavli Institute for Theoretical Physics (KITP).
KW acknowledge SPP 1992, a Priority Program funded by the Deutsche Forschungsgemeinschaft (DFG), for the support during the visit to Heidelberg, Germany in March and April 2024.

\end{acknowledgements}

%
%
\bibliographystyle{aa}
\bibliography{Qi_ffp1}

\begin{thebibliography}{94}
\expandafter\ifx\csname natexlab\endcsname\relax\def\natexlab#1{#1}\fi

\bibitem[{Adamo {et~al.}(2020)Adamo, Kruijssen, Bastian, {et~al.}}]{Adamo2020_StarClustersNearAndFar}
Adamo, A., Kruijssen, J. M.~D., Bastian, N., {et~al.} 2020, Space Science Reviews, 216, AuthorScript:51

\bibitem[{Allison {et~al.}(2009)Allison, Goodwin, Parker, Portegies~Zwart, de~Grijs, \& Kouwenhoven}]{Allison2009_MST}
Allison, R.~J., Goodwin, S.~P., Parker, R.~J., {et~al.} 2009, Monthly Notices of the Royal Astronomical Society, 395, 1449

\bibitem[{Baumgardt {et~al.}(2022)}]{Baumgardt2022_MSGlobulars}
Baumgardt, H. {et~al.} 2022, Monthly Notices of the Royal Astronomical Society, 510, 3531

\bibitem[{{Bazyey} \& {Bazyey}(2022)}]{bazyey2022}
{Bazyey}, O. \& {Bazyey}, N. 2022, Astronomical and Astrophysical Transactions, 33, 5

\bibitem[{{Belczynski} {et~al.}(2007){Belczynski}, {Taam}, {Kalogera}, {Rasio}, \& {Bulik}}]{Belczynski:2007aa}
{Belczynski}, K., {Taam}, R.~E., {Kalogera}, V., {Rasio}, F.~A., \& {Bulik}, T. 2007, \apj, 662, 504

\bibitem[{{Biermann}(1978)}]{Biermann:1978aa}
{Biermann}, L. 1978, in Astronomical Papers Dedicated to Bengt Stromgren, ed. A.~{Reiz} \& T.~{Andersen}, 327--336

\bibitem[{{Bodman} \& {Quillen}(2016)}]{Bodman:2016aa}
{Bodman}, E.~H.~L. \& {Quillen}, A. 2016, \apjl, 819, L34

\bibitem[{{Boyajian} {et~al.}(2016){Boyajian}, {LaCourse}, {Rappaport}, {Fabrycky}, {Fischer}, {Gandolfi}, {Kennedy}, {Korhonen}, {Liu}, {Moor}, {Olah}, {Vida}, {Wyatt}, {Best}, {Brewer}, {Ciesla}, {Cs{\'a}k}, {Deeg}, {Dupuy}, {Handler}, {Heng}, {Howell}, {Ishikawa}, {Kov{\'a}cs}, {Kozakis}, {Kriskovics}, {Lehtinen}, {Lintott}, {Lynn}, {Nespral}, {Nikbakhsh}, {Schawinski}, {Schmitt}, {Smith}, {Szabo}, {Szabo}, {Viuho}, {Wang}, {Weiksnar}, {Bosch}, {Connors}, {Goodman}, {Green}, {Hoekstra}, {Jebson}, {Jek}, {Omohundro}, {Schwengeler}, \& {Szewczyk}}]{Boyajian:2016aa}
{Boyajian}, T.~S., {LaCourse}, D.~M., {Rappaport}, S.~A., {et~al.} 2016, \mnras, 457, 3988

\bibitem[{{Brasser}(2008)}]{Brasser:2008ac}
{Brasser}, R. 2008, \aap, 492, 251

\bibitem[{{Brasser} {et~al.}(2008){Brasser}, {Duncan}, \& {Levison}}]{Brasser:2008ab}
{Brasser}, R., {Duncan}, M.~J., \& {Levison}, H.~F. 2008, \icarus, 196, 274

\bibitem[{{Brasser} \& {Schwamb}(2015)}]{Brasser:2015aa}
{Brasser}, R. \& {Schwamb}, M.~E. 2015, \mnras, 446, 3788

\bibitem[{{Cai} {et~al.}(2017){Cai}, {Kouwenhoven}, {Portegies Zwart}, \& {Spurzem}}]{Cai:2017aa}
{Cai}, M.~X., {Kouwenhoven}, M.~B.~N., {Portegies Zwart}, S.~F., \& {Spurzem}, R. 2017, \mnras, 470, 4337

\bibitem[{{Cai} {et~al.}(2015){Cai}, {Meiron}, {Kouwenhoven}, {Assmann}, \& {Spurzem}}]{Cai:2015aa}
{Cai}, M.~X., {Meiron}, Y., {Kouwenhoven}, M.~B.~N., {Assmann}, P., \& {Spurzem}, R. 2015, \apjs, 219, 31

\bibitem[{{Cai} {et~al.}(2019){Cai}, {Portegies Zwart}, {Kouwenhoven}, \& {Spurzem}}]{Cai:2019aa}
{Cai}, M.~X., {Portegies Zwart}, S., {Kouwenhoven}, M.~B.~N., \& {Spurzem}, R. 2019, \mnras, 489, 4311

\bibitem[{{Cai} {et~al.}(2016){Cai}, {Spurzem}, \& {Kouwenhoven}}]{Cai:2016aa}
{Cai}, M.~X., {Spurzem}, R., \& {Kouwenhoven}, M.~B.~N. 2016, in IAU Symposium, Vol. 312, Star Clusters and Black Holes in Galaxies across Cosmic Time, ed. Y.~{Meiron}, S.~{Li}, F.-K. {Liu}, \& R.~{Spurzem}, 235--236

\bibitem[{De~Vita {et~al.}(2019)}]{DeVita2019_CorrSegStruct}
De~Vita, A. {et~al.} 2019, Monthly Notices of the Royal Astronomical Society, 485, 5752

\bibitem[{{Dones} {et~al.}(2015){Dones}, {Brasser}, {Kaib}, \& {Rickman}}]{Dones:2015aa}
{Dones}, L., {Brasser}, R., {Kaib}, N., \& {Rickman}, H. 2015, \ssr, 197, 191

\bibitem[{{Dones} {et~al.}(2004){Dones}, {Weissman}, {Levison}, \& {Duncan}}]{Dones:2004aa}
{Dones}, L., {Weissman}, P.~R., {Levison}, H.~F., \& {Duncan}, M.~J. 2004, in Astronomical Society of the Pacific Conference Series, Vol. 323, Star Formation in the Interstellar Medium: In Honor of David Hollenbach, ed. D.~{Johnstone}, F.~C. {Adams}, D.~N.~C. {Lin}, D.~A. {Neufeeld}, \& E.~C. {Ostriker}, 371

\bibitem[{{Dybczy{\'n}ski}(2002)}]{Dybczynski:2002aa}
{Dybczy{\'n}ski}, P.~A. 2002, \aap, 396, 283

\bibitem[{{Eggers} {et~al.}(1997){Eggers}, {Keller}, {Kroupa}, \& {Markiewicz}}]{Eggers:1997aa}
{Eggers}, S., {Keller}, H.~U., {Kroupa}, P., \& {Markiewicz}, W.~J. 1997, \planss, 45, 1099

\bibitem[{{Eggleton} {et~al.}(1989){Eggleton}, {Fitchett}, \& {Tout}}]{Eggleton:1989aa}
{Eggleton}, P.~P., {Fitchett}, M.~J., \& {Tout}, C.~A. 1989, \apj, 347, 998

\bibitem[{{Everhart}(1967{\natexlab{a}})}]{Everhart:1967aa}
{Everhart}, E. 1967{\natexlab{a}}, \aj, 72, 716

\bibitem[{{Everhart}(1967{\natexlab{b}})}]{Everhart:1967ab}
{Everhart}, E. 1967{\natexlab{b}}, \aj, 72, 1002

\bibitem[{{Flammini Dotti} {et~al.}(2018){Flammini Dotti}, Cai, Spurzem, \& Kouwenhoven}]{dotti_cai_spurzem_kouwenhoven_2018}
{Flammini Dotti}, F., Cai, M.~X., Spurzem, R., \& Kouwenhoven, M. 2018, Proceedings of the International Astronomical Union, 14, 293

\bibitem[{{Flammini Dotti} {et~al.}(2023){Flammini Dotti}, {Capuzzo-Dolcetta}, \& {Kouwenhoven}}]{flamminidotti2023}
{Flammini Dotti}, F., {Capuzzo-Dolcetta}, R., \& {Kouwenhoven}, M.~B.~N. 2023, \mnras, 526, 1987

\bibitem[{{Flammini Dotti} {et~al.}(2025){Flammini Dotti}, {Kouwenhoven}, {Berczik}, {Shu}, \& {Spurzem}}]{flamminidotti2025}
{Flammini Dotti}, F., {Kouwenhoven}, M.~B.~N., {Berczik}, P., {Shu}, Q., \& {Spurzem}, R. 2025, \aap, 693, A166

\bibitem[{{Flammini Dotti} {et~al.}(2019){Flammini Dotti}, {Kouwenhoven}, {Cai}, \& {Spurzem}}]{2019MNRAS.489.2280F}
{Flammini Dotti}, F., {Kouwenhoven}, M.~B.~N., {Cai}, M.~X., \& {Spurzem}, R. 2019, \mnras, 489, 2280

\bibitem[{{Flammini Dotti} {et~al.}(2020){Flammini Dotti}, {Kouwenhoven}, {Shu}, {Hao}, \& {Spurzem}}]{flamminiblack}
{Flammini Dotti}, F., {Kouwenhoven}, M.~B.~N., {Shu}, Q., {Hao}, W., \& {Spurzem}, R. 2020, \mnras, 497, 3623

\bibitem[{{Fouchard} {et~al.}(2013){Fouchard}, {Rickman}, {Froeschl{\'e}}, \& {Valsecchi}}]{Fouchard:2013aa}
{Fouchard}, M., {Rickman}, H., {Froeschl{\'e}}, C., \& {Valsecchi}, G.~B. 2013, \icarus, 222, 20

\bibitem[{{Francis}(2005)}]{Francis:2005aa}
{Francis}, P.~J. 2005, \apj, 635, 1348

\bibitem[{Fregeau {et~al.}(2001)Fregeau, Joshi, Portegies~Zwart, \& Rasio}]{Fregeau2001_LightTracers}
Fregeau, J.~M., Joshi, K.~J., Portegies~Zwart, S.~F., \& Rasio, F.~A. 2001, Astrophysical Journal, 562, L5

\bibitem[{{Fujii} \& {Hori}(2019)}]{2019A&A...624A.110F}
{Fujii}, M.~S. \& {Hori}, Y. 2019, \aap, 624, A110

\bibitem[{{Giersz} \& {Spurzem}(1994)}]{Giersz:1994aa}
{Giersz}, M. \& {Spurzem}, R. 1994, \mnras, 269 [\eprint{astro-ph/9305033}]

\bibitem[{{Gomez-Leyton} \& {Velazquez}(2019)}]{gomezleyton2019}
{Gomez-Leyton}, Y.~J. \& {Velazquez}, L. 2019, \mnras, 488, 362

\bibitem[{{Hands} {et~al.}(2019){Hands}, {Dehnen}, {Gration}, {Stadel}, \& {Moore}}]{2019MNRAS.490...21H}
{Hands}, T.~O., {Dehnen}, W., {Gration}, A., {Stadel}, J., \& {Moore}, B. 2019, \mnras, 490, 21

\bibitem[{{Hao} {et~al.}(2013){Hao}, {Kouwenhoven}, \& {Spurzem}}]{Hao:2013aa}
{Hao}, W., {Kouwenhoven}, M.~B.~N., \& {Spurzem}, R. 2013, \mnras, 433, 867

\bibitem[{{Hurley} {et~al.}(2005){Hurley}, {Pols}, {Aarseth}, \& {Tout}}]{Hurley:2005aa}
{Hurley}, J.~R., {Pols}, O.~R., {Aarseth}, S.~J., \& {Tout}, C.~A. 2005, \mnras, 363, 293

\bibitem[{{Hurley} {et~al.}(2000){Hurley}, {Pols}, \& {Tout}}]{Hurley:2000aa}
{Hurley}, J.~R., {Pols}, O.~R., \& {Tout}, C.~A. 2000, \mnras, 315, 543

\bibitem[{{Hurley} {et~al.}(2002){Hurley}, {Tout}, \& {Pols}}]{Hurley:2002ab}
{Hurley}, J.~R., {Tout}, C.~A., \& {Pols}, O.~R. 2002, \mnras, 329, 897

\bibitem[{{Kamlah} {et~al.}(2022){Kamlah}, {Leveque}, {Spurzem}, {Arca Sedda}, {Askar}, {Banerjee}, {Berczik}, {Giersz}, {Hurley}, {Belloni}, {K{\"u}hmichel}, \& {Wang}}]{kamlah2022a}
{Kamlah}, A., {Leveque}, A., {Spurzem}, R., {et~al.} 2022, \mnras, 511, 4060

\bibitem[{{Kokaia} {et~al.}(2020){Kokaia}, {Davies}, \& {Mustill}}]{Kokaia2020}
{Kokaia}, G., {Davies}, M.~B., \& {Mustill}, A.~J. 2020, arXiv e-prints, arXiv:2010.15448

\bibitem[{{Kouwenhoven} {et~al.}(2010){Kouwenhoven}, {Goodwin}, {Parker}, {Davies}, {Malmberg}, \& {Kroupa}}]{Kouwenhoven:2010aa}
{Kouwenhoven}, M.~B.~N., {Goodwin}, S.~P., {Parker}, R.~J., {et~al.} 2010, \mnras, 404, 1835

\bibitem[{{Kroupa}(2001)}]{2001MNRAS.322..231K}
{Kroupa}, P. 2001, \mnras, 322, 231

\bibitem[{K{\"u}pper {et~al.}(2011)K{\"u}pper, Maschberger, Kroupa, \& Baumgardt}]{kupper}
K{\"u}pper, A.~H., Maschberger, T., Kroupa, P., \& Baumgardt, H. 2011, MNRAS, 417, 2300

\bibitem[{{Lamers} {et~al.}(2005){Lamers}, {Gieles}, {Bastian}, {Baumgardt}, {Kharchenko}, \& {Portegies Zwart}}]{Lamers:2005aa}
{Lamers}, H.~J.~G.~L.~M., {Gieles}, M., {Bastian}, N., {et~al.} 2005, \aap, 441, 117

\bibitem[{{Leveque} {et~al.}(2021){Leveque}, {Giersz}, \& {Paolillo}}]{agostino}
{Leveque}, A., {Giersz}, M., \& {Paolillo}, M. 2021, \mnras, 501, 5212

\bibitem[{{Malmberg} {et~al.}(2007){Malmberg}, {de Angeli}, {Davies}, {Church}, {Mackey}, \& {Wilkinson}}]{Malmberg:2007aa}
{Malmberg}, D., {de Angeli}, F., {Davies}, M.~B., {et~al.} 2007, \mnras, 378, 1207

\bibitem[{{Miret-Roig}(2023)}]{miretroig2023}
{Miret-Roig}, N. 2023, \apss, 368, 17

\bibitem[{{Moeckel} \& {Clarke}(2011)}]{Moeckel:2011aa}
{Moeckel}, N. \& {Clarke}, C.~J. 2011, \mnras, 415, 1179

\bibitem[{{Morbidelli}(2005)}]{Morbidelli:2005aa}
{Morbidelli}, A. 2005, ArXiv Astrophysics e-prints [\eprint{astro-ph/0512256}]

\bibitem[{{Mroz} {et~al.}(2017){Mroz}, {Ryu}, {Skowron}, {Udalski}, {Gould}, {Szymanski}, {Soszynski}, {Poleski}, {Pietrukowicz}, {Kozlowski}, {Pawlak}, {Ulaczyk}, {Albrow}, {Chung}, {Jung}, {Han}, {Hwang}, {Shin}, {Yee}, {Zhu}, {Cha}, {Kim}, {Kim}, {Kim}, {Lee}, {Lee}, {Lee}, {Park}, \& {Pogge}}]{Mroz:2017aa}
{Mroz}, P., {Ryu}, Y.-H., {Skowron}, J., {et~al.} 2017, ArXiv e-prints [\eprint[arXiv]{1712.01042}]

\bibitem[{{Offner} {et~al.}(2023){Offner}, {Moe}, {Kratter}, {Sadavoy}, {Jensen}, \& {Tobin}}]{offner2023}
{Offner}, S.~S.~R., {Moe}, M., {Kratter}, K.~M., {et~al.} 2023, in Astronomical Society of the Pacific Conference Series, Vol. 534, Protostars and Planets VII, ed. S.~{Inutsuka}, Y.~{Aikawa}, T.~{Muto}, K.~{Tomida}, \& M.~{Tamura}, 275

\bibitem[{Olczak {et~al.}(2011)Olczak, Spurzem, \& Henning}]{Olczak2011_effMST}
Olczak, C., Spurzem, R., \& Henning, T. 2011, Astronomy \& Astrophysics, 532, A119

\bibitem[{{Oort}(1950)}]{Oort:1950aa}
{Oort}, J.~H. 1950, \bain, 11, 91

\bibitem[{{Parker} {et~al.}(2016){Parker}, {Goodwin}, {Wright}, {Meyer}, \& {Quanz}}]{parker2016}
{Parker}, R.~J., {Goodwin}, S.~P., {Wright}, N.~J., {Meyer}, M.~R., \& {Quanz}, S.~P. 2016, \mnras, 459, L119

\bibitem[{{Parker} {et~al.}(2017){Parker}, {Lichtenberg}, \& {Quanz}}]{Parker:2017aa}
{Parker}, R.~J., {Lichtenberg}, T., \& {Quanz}, S.~P. 2017, \mnras, 472, L75

\bibitem[{{Parker} \& {Quanz}(2012)}]{parker2012}
{Parker}, R.~J. \& {Quanz}, S.~P. 2012, \mnras, 419, 2448

\bibitem[{{Pe{\~n}arrubia}(2023)}]{pena2023}
{Pe{\~n}arrubia}, J. 2023, \mnras, 519, 1955

\bibitem[{{Pelupessy} {et~al.}(2013){Pelupessy}, {van Elteren}, {de Vries}, {McMillan}, {Drost}, \& {Portegies Zwart}}]{Pelupessy:2013aa}
{Pelupessy}, F.~I., {van Elteren}, A., {de Vries}, N., {et~al.} 2013, \aap, 557, A84

\bibitem[{{Perets} \& {Kouwenhoven}(2012)}]{Perets:2012aa}
{Perets}, H.~B. \& {Kouwenhoven}, M.~B.~N. 2012, \apj, 750, 83

\bibitem[{{Pfalzner} \& {Bannister}(2019)}]{Pfalzner2019}
{Pfalzner}, S. \& {Bannister}, M.~T. 2019, \apjl, 874, L34

\bibitem[{{Plummer}(1911)}]{Plummer:1911aa}
{Plummer}, H.~C. 1911, \mnras, 71, 460

\bibitem[{{Portegies Zwart}(2020)}]{Portegies:2020aa}
{Portegies Zwart}, S. 2020, arXiv e-prints, arXiv:2011.08257

\bibitem[{{Portegies Zwart}(2021)}]{pz2021a}
{Portegies Zwart}, S. 2021, \aap, 647, A136

\bibitem[{{Portegies Zwart} {et~al.}(2013){Portegies Zwart}, {McMillan}, {van Elteren}, {Pelupessy}, \& {de Vries}}]{Portegies-Zwart:2013aa}
{Portegies Zwart}, S., {McMillan}, S.~L.~W., {van Elteren}, E., {Pelupessy}, I., \& {de Vries}, N. 2013, Computer Physics Communications, 183, 456

\bibitem[{{Portegies Zwart} {et~al.}(2021){Portegies Zwart}, {Torres}, {Cai}, \& {Brown}}]{pz2021b}
{Portegies Zwart}, S., {Torres}, S., {Cai}, M.~X., \& {Brown}, A. G.~A. 2021, \aap, 652, A144

\bibitem[{{Raymond} {et~al.}(2024){Raymond}, {Kaib}, {Selsis}, \& {Bouy}}]{raymond2024}
{Raymond}, S.~N., {Kaib}, N.~A., {Selsis}, F., \& {Bouy}, H. 2024, \mnras, 527, 6126

\bibitem[{{Sana} {et~al.}(2012){Sana}, {Dunstall}, {H{\'e}nault-Brunet}, {Walborn}, {de Koter}, {de Mink}, {Dufton}, {Evans}, {Ma{\'\i}z Apell{\'a}niz}, {Taylor}, \& {Vink}}]{sana2012}
{Sana}, H., {Dunstall}, P.~R., {H{\'e}nault-Brunet}, V., {et~al.} 2012, in Astronomical Society of the Pacific Conference Series, Vol. 465, Proceedings of a Scientific Meeting in Honor of Anthony F. J. Moffat, ed. L.~{Drissen}, C.~{Robert}, N.~{St-Louis}, \& A.~F.~J. {Moffat}, 284

\bibitem[{{Seligman} \& {Moro-Mart{\'\i}n}(2022)}]{seligman2022}
{Seligman}, D.~Z. \& {Moro-Mart{\'\i}n}, A. 2022, Contemporary Physics, 63, 200

\bibitem[{{Shannon} {et~al.}(2015){Shannon}, {Jackson}, {Veras}, \& {Wyatt}}]{Shannon:2015aa}
{Shannon}, A., {Jackson}, A.~P., {Veras}, D., \& {Wyatt}, M. 2015, \mnras, 446, 2059

\bibitem[{{Shara} {et~al.}(2016){Shara}, {Hurley}, \& {Mardling}}]{Shara:2016aa}
{Shara}, M.~M., {Hurley}, J.~R., \& {Mardling}, R.~A. 2016, \apj, 816, 59

\bibitem[{{Shu} {et~al.}(2020){Shu}, {Pang}, {Flammini Dotti}, {Kouwenhoven}, {Arca Sedda}, \& {Spurzem}}]{Shu2020}
{Shu}, Q., {Pang}, X., {Flammini Dotti}, F., {et~al.} 2020, arXiv e-prints, arXiv:2011.14911

\bibitem[{{Siraj} \& {Loeb}(2019)}]{Siraj:2018aa}
{Siraj}, A. \& {Loeb}, A. 2019, \apjl, 872, L10

\bibitem[{{Siraj} \& {Loeb}(2020)}]{Siraj:2020aa}
{Siraj}, A. \& {Loeb}, A. 2020, arXiv e-prints, arXiv:2011.14900

\bibitem[{{Spitzer}(1987)}]{Spitzer:1987aa}
{Spitzer}, L. 1987, {Dynamical evolution of globular clusters} (Princeton University Press)

\bibitem[{{Spurzem} {et~al.}(2009){Spurzem}, {Giersz}, {Heggie}, \& {Lin}}]{Spurzem:2009aa}
{Spurzem}, R., {Giersz}, M., {Heggie}, D.~C., \& {Lin}, D.~N.~C. 2009, \apj, 697, 458

\bibitem[{{Spurzem} \& {Kamlah}(2023)}]{spurzem2023}
{Spurzem}, R. \& {Kamlah}, A. 2023, Living Reviews in Computational Astrophysics, 9, 3

\bibitem[{{Stock} {et~al.}(2020){Stock}, {Cai}, {Spurzem}, {Kouwenhoven}, \& {Portegies Zwart}}]{stock2020}
{Stock}, K., {Cai}, M.~X., {Spurzem}, R., {Kouwenhoven}, M.~B.~N., \& {Portegies Zwart}, S. 2020, \mnras, 497, 1807

\bibitem[{{Torbett}(1986)}]{Torbett:1986aa}
{Torbett}, M.~V. 1986, \aj, 92, 171

\bibitem[{{Torres} {et~al.}(2019){Torres}, {Cai}, {Brown}, \& {Portegies Zwart}}]{Torres:2019aa}
{Torres}, S., {Cai}, M.~X., {Brown}, A.~G.~A., \& {Portegies Zwart}, S. 2019, arXiv e-prints [\eprint[arXiv]{1906.10617}]

\bibitem[{{Valtonen} \& {Innanen}(1982)}]{Valtonen:1982aa}
{Valtonen}, M.~J. \& {Innanen}, K.~A. 1982, \apj, 255, 307

\bibitem[{{Veras} {et~al.}(2020){Veras}, {Reichert}, {Flammini Dotti}, {Cai}, {Mustill}, {Shannon}, {McDonald}, {Portegies Zwart}, {Kouwenhoven}, \& {Spurzem}}]{Veras2019}
{Veras}, D., {Reichert}, K., {Flammini Dotti}, F., {et~al.} 2020, \mnras, 493, 5062

\bibitem[{{Veras} \& {Wyatt}(2012)}]{Veras:2012aa}
{Veras}, D. \& {Wyatt}, M.~C. 2012, \mnras, 421, 2969

\bibitem[{{Wang} {et~al.}(2015){Wang}, {Kouwenhoven}, {Zheng}, {Church}, \& {Davies}}]{Wang:2015ab}
{Wang}, L., {Kouwenhoven}, M.~B.~N., {Zheng}, X., {Church}, R.~P., \& {Davies}, M.~B. 2015, \mnras, 449, 3543

\bibitem[{{Wang} {et~al.}(2020){Wang}, {Perna}, \& {Leigh}}]{wangyihan}
{Wang}, Y.-H., {Perna}, R., \& {Leigh}, N. W.~C. 2020, \mnras, 496, 1453

\bibitem[{{Weissman}(1980)}]{Weissman:1980aa}
{Weissman}, P.~R. 1980, \nat, 288, 242

\bibitem[{{Weissman}(1982)}]{Weissman:1982aa}
{Weissman}, P.~R. 1982, in IAU Colloq. 61: Comet Discoveries, Statistics, and Observational Selection, ed. L.~L. {Wilkening}, 637--658

\bibitem[{{Weissman}(1996)}]{Weissman:1996aa}
{Weissman}, P.~R. 1996, in Astronomical Society of the Pacific Conference Series, Vol. 107, Completing the Inventory of the Solar System, ed. T.~{Rettig} \& J.~M. {Hahn}, 265--288

\bibitem[{{Welsh} \& {Montgomery}(2016)}]{Welsh:2016aa}
{Welsh}, B.~Y. \& {Montgomery}, S. 2016, \pasp, 128, 064201

\bibitem[{{Wiegert} \& {Tremaine}(1999)}]{Wiegert:1999aa}
{Wiegert}, P. \& {Tremaine}, S. 1999, \icarus, 137, 84

\bibitem[{{Wu} {et~al.}(2024{\natexlab{a}}){Wu}, {Kouwenhoven}, {Dotti}, \& {Spurzem}}]{wukai2024}
{Wu}, K., {Kouwenhoven}, M.~B.~N., {Dotti}, F.~F., \& {Spurzem}, R. 2024{\natexlab{a}}, \mnras

\bibitem[{{Wu} {et~al.}(2024{\natexlab{b}}){Wu}, {Kouwenhoven}, {Flammini Dotti}, \& {Spurzem}}]{kai2024}
{Wu}, K., {Kouwenhoven}, M.~B.~N., {Flammini Dotti}, F., \& {Spurzem}, R. 2024{\natexlab{b}}, \mnras, 533, 4485

\bibitem[{{Wu} {et~al.}(2023){Wu}, {Kouwenhoven}, {Spurzem}, \& {Pang}}]{wukai2023}
{Wu}, K., {Kouwenhoven}, M.~B.~N., {Spurzem}, R., \& {Pang}, X. 2023, \mnras, 523, 4801

\bibitem[{{Zheng} {et~al.}(2015){Zheng}, {Kouwenhoven}, \& {Wang}}]{Zheng:2015aa}
{Zheng}, X., {Kouwenhoven}, M.~B.~N., \& {Wang}, L. 2015, \mnras, 453, 2759

\end{thebibliography}

\begin{appendix}

\section{An alternative approach on the decoupling of the mass segregation timescale} \label{sec:msdisc}

\fff{T}he role of segregation of particles in a star cluster is mostly influenced by the stars. Small mass stars are segregated outside, and heavy masses go in the center. However, there is a certain upper limit of mass for which this no longer applies. In this work, we try to analyse and find this limit using two different approaches of calculating the mass segregation timescale. However, in this section we will first discuss our formalism, using Section \ref{sec:MSTD} for an analytical discussion.
For clarity purposes, we will describe the low mass particles population as LMP and the stars as HMP (high mass particles population)
The mass segregation timescale for a system with two populations can be written as
\begin{equation}
    t_{\rm ms} = \frac{\langle m_{\rm av, LMP \& HMP} \rangle}{m_{\rm max}} t_{\rm rh} 
    \quad , 
\end{equation}
where $\langle m_{\rm av,LMP \& HMP} \rangle$ 
is the average stellar mass considering both populations, and $m_{\rm max}$ is the maximum HMP mass (as $m_{\rm max,HMP} \gg  m_{\rm max,LMP} $ by definition). This expression can be used to obtain estimates for the timescale at which mass segregation occurs in a typical star cluster. The above expression, however, does not provide information about the behavior of a LMP, that may be present in a star cluster. 

We can rewrite the total average mass as:
\begin{equation}
    \langle m_{\rm av, LMP \& HMP} \rangle = \frac{N_{\rm LMP} \ m_{\rm LMP}}{N_{\rm LMP}+N_{\rm HMP}} + \frac{M_{\rm cl}}{N_{\rm LMP}+N_{\rm HMP}} \ \quad,
\end{equation}
where $N_{\rm LMP}$ and $N_{\rm HMP}$ are the number of components of the LMP and HMP respectively, while $m_{\rm LMP}$ is the mass of a single member of the LMP. 
Since we chose equal-mass LMP in our previously shown simulations, we treat the total mass of the LMP population as $N_{\rm LMP} \ m_{\rm LMP}$ (using a mass spectrum for the LMP would not change this result drastically, thus we use a same mass distribution). 
If we assume that the two populations are not decoupled (and thus the HMP gravitationally control the dynamical evolution of the LMP of mass segregation), then we can assume that the HMP relaxation time dominates. 
The mass segregation timescale is the same of the entire system:
\begin{equation}
    t_{\rm ms,LMP,1} = t_{\rm rh} \frac{N_{\rm LMP} \ m_{\rm LMP}}{m_{\rm max}} \ \quad,
    \label{eq_tms_ffc_gen}
\end{equation}
where $t_{\rm ms} = t_{\rm ms,HMP} + t_{\rm ms, LMP}$ (as they have $t_{\rm rh}$ as common parameter). 
Notice that we are making assumptions only on the average mass of the two populations, so it is independent from the number and mass of the LMP objects compared to the HMP ones.
The timescale can be generalized for any population of LMP. The maximum mass is the HMP largest mass.
We can also write the equation showing the significant physical quantities, resulting in:
\begin{equation}
    t_{\rm ms,LMP1} = k N_{\rm LMP} \ln(0.02(N_{\rm LMP}+N_{\rm HMP})) \Delta m \ m_{\rm max}^{-1} r_{\rm hm}^{3/2} \quad,
    \label{eq_tms_ffc_gen_1}
\end{equation}
%
where $\Delta m = m_{\rm LMP} / M_{\rm cl}^{1/2}$  and $k = 0.138 G^{-1/2}$. Since it is a log value multiplied by 0.02, $N_{\rm tot}$ has at best one order of magnitude for large value of the total number of objects. Moreover, this is a term which is multiplied for both mass segregation timescales.\\
Using the initial condition of our star cluster model, excluding the LMP initial conditions, we obtain:
\begin{equation}
    t_{\rm ms,LMP1} = 2.95 \times 10^{-4} {\rm Myr} \frac{N_{\rm LMP}}{\ln(0.02 (N_{\rm HMP} + N_{\rm LMP}))} \frac{m_{\rm LMP}}{\msun} \quad,
    \label{eq_tms_ffc_gen_1m}
\end{equation}
where we left $N_{\rm HMP}$ implicit for consistency.\\
We use a similar logic for decoupled populations. The total number of stars and the total mass of the system are changed to $N_{\rm LMP}$ and $N_{\rm LMP} \ m_{\rm LMP}$, respectively. We can treat the terms of mass segregation for the population completely separated. Moreover, the relaxation time of the populations is treated separately.
Thus, Equation~\ref{eq_tms_ffc_gen} becomes:
\begin{equation}
    t_{\rm ms,LMP2} = k N_{\rm LMP}^{3/2} m_{\rm LMP}^{-1/2} N_{\rm LMP \& HMP}^{-1}r_{\rm hm}^{3/2} \ln(0.11(N_{\rm LMP})) \quad,
    \label{eq_tms_ffc_gen_2}
\end{equation}
where we have also a dependence on the total number of particles of the system, which is negligible when $N_{\rm LMP} \gg N_{\rm HMP}$.
Using the initial conditions of the star clusters of our previously shown simulations but excluding the number of LMP and the mass of the LMP, we have:
\begin{equation}
    t_{\rm ms,LMP2} = 2.16 \ {\rm Myr} \frac{N_{\rm LMP}}{\ln(0.22 N_{\rm LMP}) } \Big(\frac{m_{\rm LMP}}{\msun}\Big)^{-1/2} \quad,
    \label{eq_tms_ffc_gen_2m}
\end{equation}

Equations \ref{eq_tms_ffc_gen_1m} and \ref{eq_tms_ffc_gen_2m} depends solely on the mass and the number of the LMP objects. In the majority of models regarding LMP, $N_{\rm HMP} \ll N_{\rm LMP}$, thus the main difference is related to the density distribution of the HMP and the number distribution of the LMP. 
If we compare the two timescale in Equation~\ref{eq_tms_ffc_gen_1} and ~\ref{eq_tms_ffc_gen_2} by dividing them, we find that, for $N_{\rm LMP} \gg N_{\rm HMP}$:

\begin{equation}
    \frac{t_{\rm ms,LMP2}}{t_{\rm ms,LMP1}} = N_{\rm LMP} m_{\rm LMP}^{-3/2} M_{\rm cl} m_{\rm max}
\end{equation}\label{eq:ratioequs}

where we find that there is a strong dependence on the cluster mass, the initial mass function and the characteristic mass and number of objects of the LMP population.


\section{Core evolution}\label{app1}

\begin{figure}[h]
\begin{tabular}{c}
  \includegraphics[width=\columnwidth]{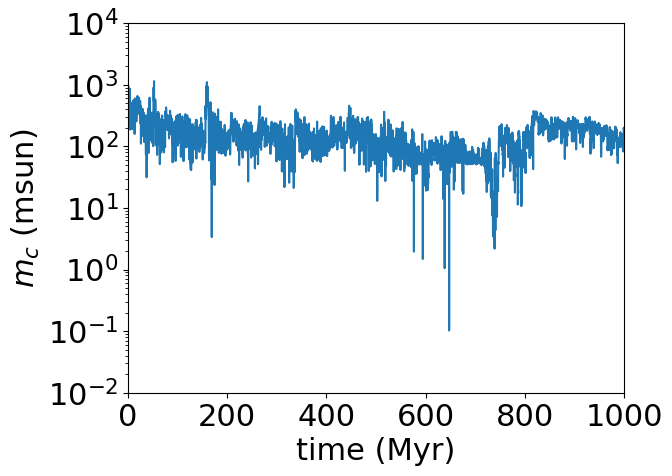}\\
  \includegraphics[width=\columnwidth]{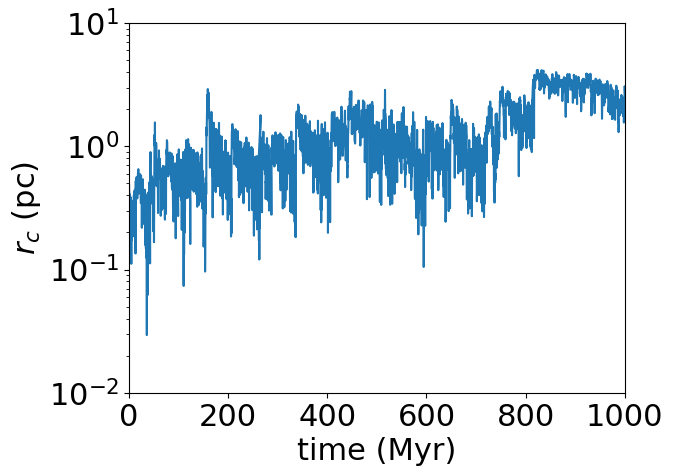}\\
\end{tabular}
  \caption{Core radius (bottom) and core mass (top) for the stellar population of model C05. 
  }\label{fig:corevol}
\end{figure}


\newpage
\newpage

\section{Single model distribution of velocity and position}\label{app2}

\begin{figure}[h]
\begin{tabular}{c}
  \includegraphics[width=\columnwidth]{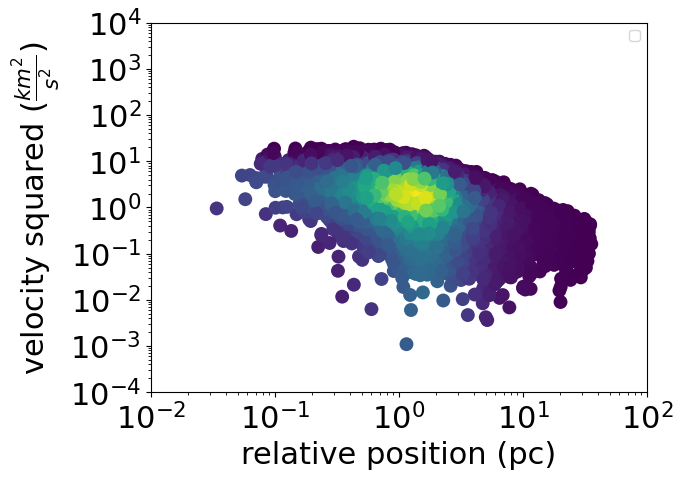}\\
  \includegraphics[width=\columnwidth]{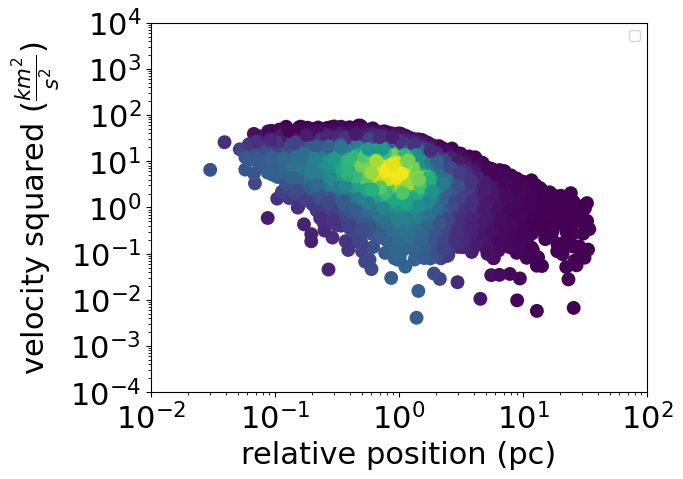}\\
  \includegraphics[width=\columnwidth]{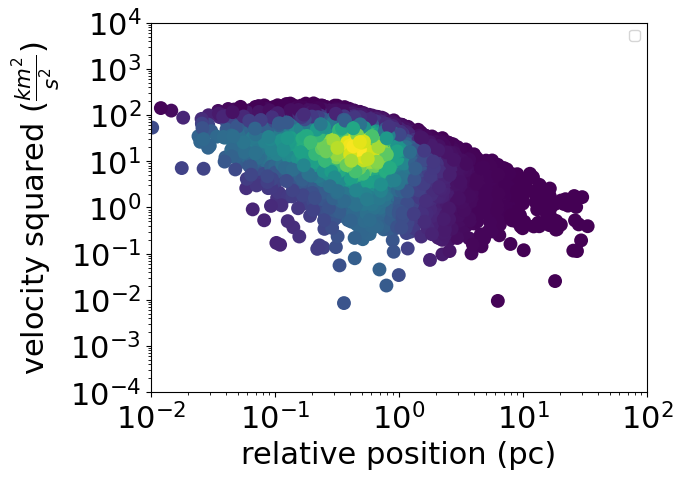}\\
\end{tabular}
  \caption{Distribution of velocity square and relative position \fff{of \ffc{}s}  at $t=0$ for models C025 (top), C05 (middle), and C075 (bottom).
  }\label{fig:startsinglemodel}
\end{figure}

\begin{figure}[h]
\begin{tabular}{c}
  \includegraphics[width=\columnwidth]{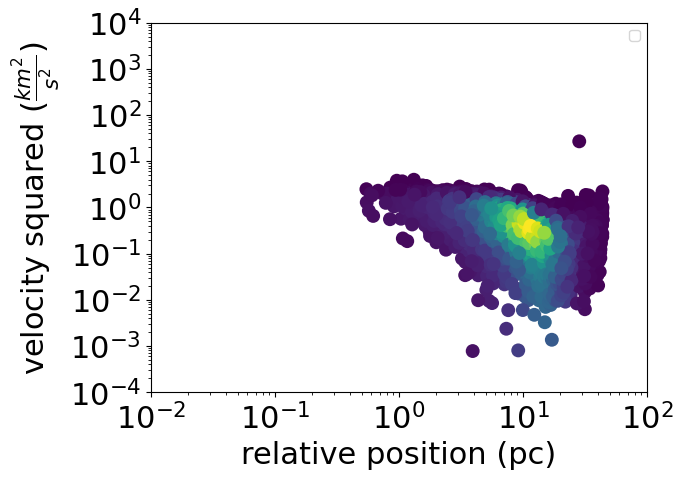}\\
  \includegraphics[width=\columnwidth]{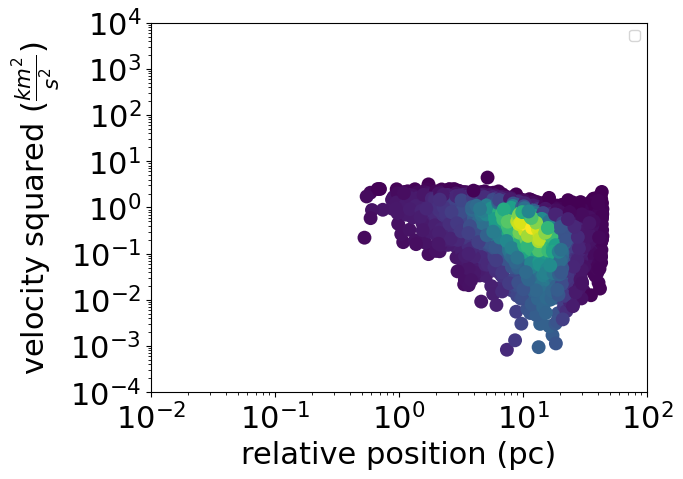}\\
  \includegraphics[width=\columnwidth]{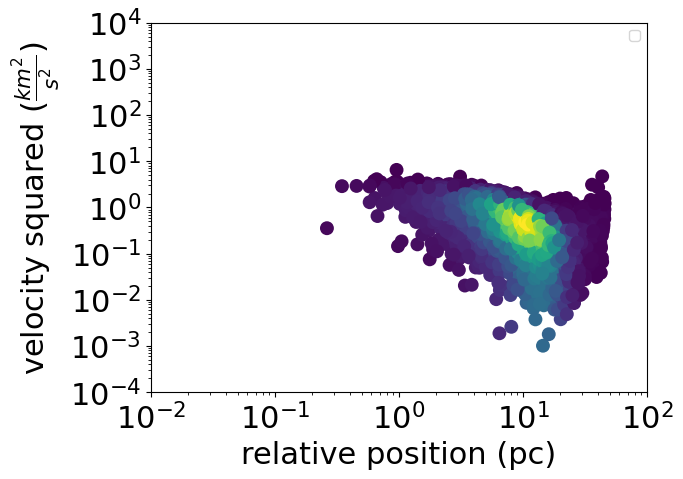}\\
\end{tabular}
  \caption{Distribution of velocity square and relative position \fff{of \ffc{}s} at 1~Gyr for models C025 (top), C05 (middle), and C075 (bottom).
  }\label{fig:endsinglemodel}
\end{figure}

\newpage
\newpage


\section{Escape velocity of stars and \ffc{}s in all models}\label{app3}

\begin{figure}[h]
\begin{tabular}{c}
  \includegraphics[width=\columnwidth]{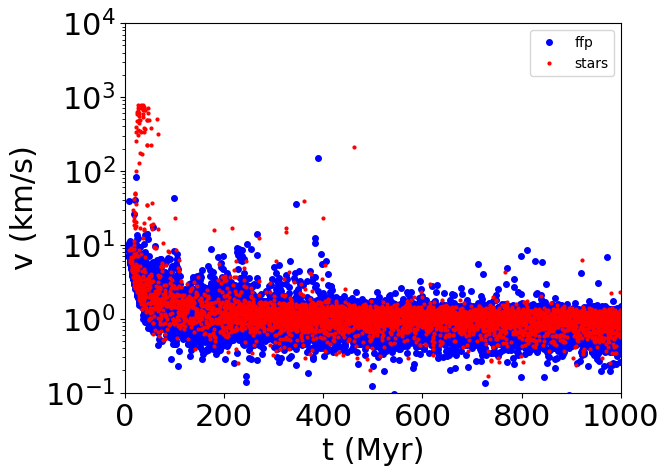}\\
  \includegraphics[width=\columnwidth]{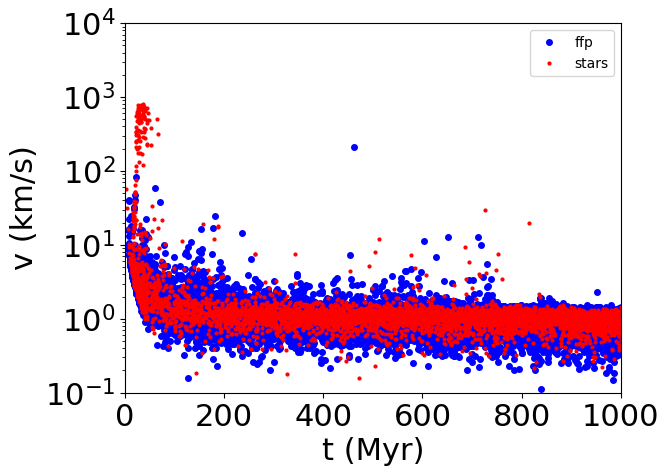}\\
  \includegraphics[width=\columnwidth]{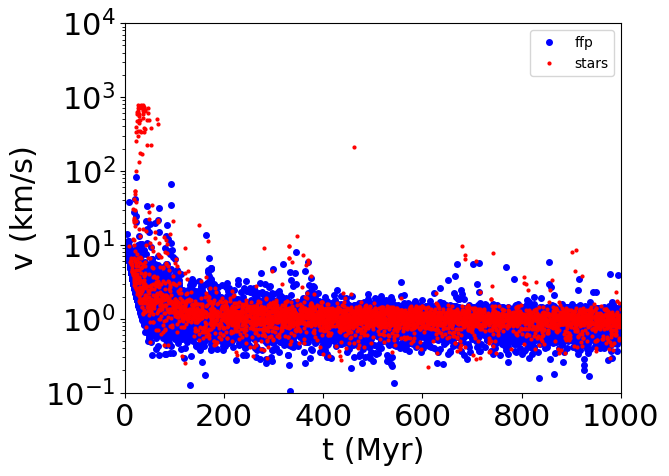}\\
\end{tabular}
  \caption{Speed of escaping \ffc{}s (blue) and stars (red) at 1~Gyr for models C025 (top), C05 (middle), and C075 (bottom). 
  }\label{fig:endsinglemodel}
\end{figure}


\section{Velocity dispersion  of \ffc{}s}\label{app4}

\begin{figure}[h]
\begin{tabular}{c}
  \includegraphics[width=\columnwidth]{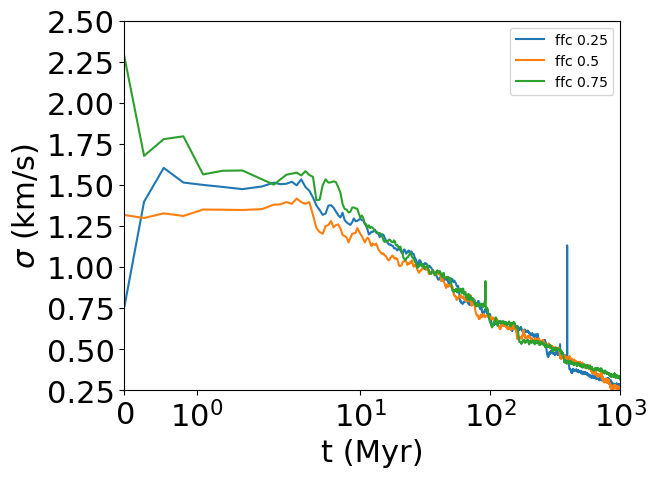}\\
\end{tabular}
  \caption{Velocity dispersion of all \ffc{} in all models. The initial imprint of the velocity distribution of \ffc{} is fastly removed in less than 1 Myr, with all models having similar velocity dispersion around 40 Myr. The spikes shown in the figure are due to high velocity \ffc{} due to a stellar encounter.
  }\label{fig:sigmaffconly}
\end{figure}

\end{appendix}

\end{document}